\documentclass{aa}  

\usepackage{graphicx}
\usepackage{txfonts}
\usepackage[colorlinks=true, linkcolor=blue, citecolor=blue]{hyperref}
\usepackage{pifont}
\usepackage{bbding}
\usepackage[table]{xcolor} 
\definecolor{softgreen}{RGB}{200, 255, 200}
\definecolor{darkgreen}{RGB}{0, 128, 0}  
\definecolor{yellow}{RGB}{255, 255, 0}       
\definecolor{darkyellow}{RGB}{204, 183, 0}   

\usepackage{natbib}
\bibpunct{(}{)}{;}{a}{}{,} 
\usepackage{longtable}
\usepackage{lscape} 
\usepackage{booktabs} 
\usepackage{multirow} 

\newcommand{\Macc}{$\dot{M}_{\mathrm{acc}}\xspace$}
\newcommand{\Lacc}{$L_{\rm acc}\xspace$}
\newcommand{\Lline}{$L_{\rm H \textsc{i}}$}

\newcommand{\Msun}{$M_{\odot}\xspace$}
\newcommand{\Lbol}{$L_{\rm bol}$}

\newcommand{\mum}{$\mu$m}
\newcommand{\hi}{H\,{\sc i}\xspace}
\newcommand{\hii}{H\,{\sc ii}\xspace}

\begin{document}

\title{Unlocking accretion rate diagnostics for high-mass protostars using JWST/MIRI  HI lines}

   \author{S. D. Reyes-Reyes  \inst{1}
          \and
          H. Beuther          \inst{1} 
          \and
          E. F. van Dishoeck  \inst{2,3}
          \and
          C. Gieser            \inst{1}
          \and
          A. Caratti o Garatti \inst{4}
          \and 
           Ł. Tychoniec        \inst{2}
           \and
           P. J. Kavanagh       \inst{5}
           \and
           P. D. Klaassen       \inst{6}
           \and
           K. Justtanont        \inst{7}
           \and
            L. Francis          \inst{2}
            \and
            V. J. M. Le Gouellec \inst{8,9,10}
            \and
           R. Devaraj           \inst{11}
           \and
           T. P. Ray            \inst{11}
          \and
          Y. Chen               \inst{2}
           \and
          M. G. Navarro               \inst{12}
           \and
           W. R. M. Rocha       \inst{2,13}
           \and
          M. L. van Gelder      \inst{2}
          }

    \institute{
    	Max-Planck-Institut für Astronomie, Königstuhl 17, D-69117 Heidelberg, Germany\\
              \email{reyes@mpia.de}
          \and
	    Leiden Observatory, Leiden University, PO Box 9513, 2300 RA Leiden, The Netherlands
        \and
         Max Planck Institut für Extraterrestrische Physik (MPE), Giessen-bachstrasse 1, 85748 Garching, Germany
        \and
         INAF–Osservatorio Astronomico di Capodimonte, Salita Moiariello 16, 80131 Napoli, Italy
         \and
          Department of Physics, Maynooth University, Maynooth, Co. Kildare, Ireland
         \and
          UK Astronomy Technology Centre, Royal Observatory Edinburgh, Blackford Hill, Edinburgh EH9 3HJ, UK
          \and
          Department of Space, Earth and Environment, Chalmers University of Technology, Onsala Space Observatory, 439 92 Onsala, Sweden
          \and
          Space Science and Astrobiology Division, NASA’s Ames Research Center, Moffett Field, CA 94035, USA
          \and
          Institut de Ciencies de l’Espai (ICE-CSIC), Campus UAB, Carrer de Can Magrans S/N, 08193 Cerdanyola del Valles, Catalonia, Spain
          \and
          Institut d’Estudis Espacials de Catalunya (IEEC), c/ Gran Capitá, 2–4, 08034 Barcelona, Spain    
          \and
          School of Cosmic Physics, Dublin Institute for Advanced Studies, 31 Fitzwilliam Place, D02 XF86, Dublin, Ireland
          \and
           INAF-Osservatorio Astronomico di Roma, Via di Frascati 33, 00078 Monte Porzio Catone, Italy
          \and
          Laboratory for Astrophysics, Leiden Observatory, Leiden University, PO Box 9513, 2300 RA Leiden, The Netherlands   
             }

   \date{Received December 15, 2025; accepted }

  \abstract
   {While many aspects of high-mass star formation have been investigated, the accretion onto the central protostars is one of the most fundamental but less explored physical properties. The James Webb Space Telescope (JWST), through its Mid InfraRed Instrument (MIRI), offers a unique opportunity to explore tracers of accretion at less-extincted wavelengths (5 to 27 $\mu$m) than those studied so far, where it delivers unparalleled sensitivity and spectral resolution.}
   {We probe the capability of MIRI in its MRS/IFU mode to detect and resolve atomic Hydrogen (H\,{\sc i}) emission lines in such young and embedded objects, to subsequently estimate accretion luminosities (\Lacc{}) and accretion rates (\Macc{}) for the first time in a sample of (six) high-mass star forming regions at different evolutionary stages.}
   {We use the dereddened \hi line luminosities as tracers of accretion by applying line-to-accretion-luminosity relations (\Lacc{}-calibrations) from literature. As such \Lacc{}-calibrations were originally established for low-mass Class II objects, we assess their applicability on our sample prior to estimating accretion rates. Extinction values were estimated from the broad silicate absorption feature at 9.7 $\mu$m.}
   {The infrared continuum reveals, at much higher spatial resolution than before, the location of new IR sources (protostars), toward which we detect a handful of \hi lines. While a few lines are secure detections, many are tentative. The most commonly detected line is the Humphreys $\alpha$ at 12.37 \mum{}, followed by Humphreys $\beta$ and Pfund $\alpha$. Assuming that their line fluxes are dominated by accretion, we find that two of the three existing \Lacc-calibrations predict excessively high accretion luminosities that largely exceed their bolometric luminosities ($L_{\rm bol}$), and that the third \Lacc{}-calibration still overpredicts accretion luminosities for some sources. Considering the given uncertainties, estimated accretion rates are only tentative.}
   {This work demonstrates the great potential of JWST/MIRI to probe \hi line emission originated in the innermost regions of high-mass protostellar systems, setting the ground floor for further investigations into accretion in these objects.  While this project had the ambitious goal of robustly quantifying accretion rates, we have shed light on what outstanding methodological challenges remain, where developing new \Lacc{}- calibrations for intermediate- to high-mass protostars appears as the most critical one.
   }

   \keywords{Accretion, accretion disks -- Stars: formation -- Stars: massive -- Stars: protostars 
				-- Stars: winds, outflows
               }

   \maketitle
%

\section{Introduction} \label{Intro}

The processes governing the accretion of material from the inner protoplanetary disk onto the central star likely vary across mass regimes. While they are well-known for low-mass pre-main sequence (PMS) stars with disks, much less is known for the intermediate and high-mass sources. On top of this, accretion during the deeply embedded phases (first $\sim 10^{5}$ years for low-mass protostars; \citealt{Dunham2015}), where the central star and disk are obscured by the dusty envelope, is poorly understood, and yet it is when stars gain the majority of their final masses (see review of \cite{Hartmann2016} for masses below 5 $M_{\odot}$, and \citealt{Beuther2025} for a comprehensive review of both low- and high- mass regimes).

For low-mass PMS stars with disks (T-Tauri Stars, also TTS), the general picture is clearer as their dissipated envelope no longer obscures the optical or infrared (IR) light coming from its central source. At this phase, the disk is truncated by the magnetic field produced by the star, while the star's radiation sublimates the inner disk dust within a dust destruction radius. As a consequence, flows of material fall onto the star due to gravity, at free fall speeds, following the magnetic field lines \citep{Koenigl1991}, while another fraction can be ejected as MHD or photoevaporative winds and jets. The inflows of material are episodic, and in the form of column or funnel flows that reach temperatures of about 10$^4$ K.  Near the stellar chromosphere a shock is produced, converting kinetic energy into radiation \citep{Bhandare2025} and briefly raising temperatures up to $\sim 10^6$ K \citep{Hartmann2016}. This radiation can be seen as a continuum excess that peaks in the UV \citep{Koenigl1991} but can also be observed at the optical and NIR.
The gas along the accretion funnels is excited and shows  \hi lines emitting at optical (e.g. H${\, \alpha}$, H${\, \beta}$ lines; \citealt{Hamann1992, Bouvier2007}), near infrared (NIR; e.g. Pa$\alpha$, Pa$\beta$, Br$\alpha$) and mid-infrared (MIR; e.g. Hu$\, \alpha$; \citealt{Beuther23}) wavelengths, both from the shock and funnel flows (and potentially winds).  Due to their larger surface area, funnel flows may dominate the bulk of the observed line emission.

For the high-mass regime ($M \geq$ 8 $ \rm M_{\odot}$) it still remains unknown whether the same (magnetospheric) processes hold. Massive sources may no longer accrete through funnel flows because they are thought to become rather radiative instead of convective as they grow in mass \citep{Stahler_Palla2004}, meaning that their magnetic field should no longer dominate the physics of accretion. It is also unclear whether the \hi emission in these objects is only tracing accretion or might be dominated by circumstellar ionized gas. 

An alternative theory of disk evolution (derived for low mass TTSs) that may explain the accretion onto massive stars is the Boundary Layer scenario presented in \citealt{Lynden-Bell1974}. In the case where magnetic fields are not dominant, the disk extends towards the stellar photosphere (filling the gap produced in the magnetospheric process) and grazes the relatively steady stellar atmosphere, producing that the material enters the star while angular momentum gets dissipated outwards. Here, emission would arise from viscous dissipation that converts part of the gravitational energy into continuum luminosity that peaks in the UV, similarly to the magnetospheric shocks, while the variability might be explained by the suddenness of the entry of orbital material into the stellar atmosphere \citep{Lynden-Bell1974}.

To produce emission lines, the Hydrogen electrons must be excited to higher quantum levels by a certain energy source such that they can spontaneously de-excite or recombine to lower levels. A couple models exist in literature to explain what processes drive this energy input. 
Case B recombination \citep{Baker1938} was initially proposed to explain the Balmer decrement on Nebulae under the hypothesis that it is opaque to the Lyman series, but it has been invoked to also explain the \hi emission lines seen in Classical T-Tauri Stars (CTTS), in which case the inflowing plasma gets photoionized, producing free electrons that can subsequently recombine. Under this assumption, the accreting flows would produce broad emission lines of atoms and molecules, while the region of the shock would generate relatively narrow lines \citep{Hartmann2016}. Until the growing central star has not yet reached 8 M$_{\odot}$, shock UV radiation might be the responsible of heating up the funnel flows till ionization. If not radiation, an unknown magnetic mechanism may apply \citep{Hartmann2016}. Case B emissivities have been studied in \citet{Storey-Hummer1987} while testing collisional excitation effects up to quantum level $n=3$. They find that if collisions from the ground states to level three are important, then derived Case B emissivity ratios are no longer valid because it implies contamination of other upper levels as well. 

More recently, alternative emissivity ratio models of \citet{Kwan2011} (KF11) that study the local conditions of the flowing material (either accretion flows or winds), suggested that collisional excitation (CE) is the primary mechanism responsible of the \hi line production. Here, collisions are powered by the intrinsic thermal kinetic energy of the flow itself, relegating recombination due to irradiation (Case B) to have a less-significant role. Collisional excitation points to flow temperatures between 8750 and 1.25$\times 10^{4}$ K, and to \hi line emissivities that strongly depend on volume densities and temperature. Both Case B and CE suggest specific Hydrogen line ratios under different conditions of temperature, electron density (Case B), volume density (CE), or ionization rate (CE), which can be used to infer the properties of the gas producing the lines depending on the adopted model. In this regard, the \hi line ratios observed by \citet{Edwards2013} for a sample of 16 TTSs give support to the local conditions scenario where collisions are protagonists. 
Given all of these, not a single theory has yet been derived to explain line emission in the environments of high-mass ($ \geq 8 M_{\odot}$) protostars. This makes sense because high-mass sources are much harder to probe since they transition directly from protostellar phase to the main sequence \citep{Palla1999}, without offering an exposed disk while it still accretes material (pre-main sequence phase). 

The most direct approach to measure accretion luminosity ($L_{\rm acc}$) is obtained from excess emission above the photosphere continuum, most commonly in the UV but also in the optical or IR \citep{Gullbring2000, Donehew2011}, and it can be studied simultaneously with \hi lines in PMS stars to search for scaling relations that allow us to estimate accretion luminosities (i.e. a "calibration") when direct methods are no longer possible, e.g. due to high extinction in the line-of-sight (LOS). In this regard, works focused on NIR lines like  Pa${\,\beta}$, Br${\,\gamma}$ or Pf${\,\beta}$ \citep{Grant2022, Muzerolle1998b, Alcala2014, Salyk2013} have found such correlations, confirming that the strength of \hi lines can be used as tracer of \Lacc{}. In protostellar systems, extinctions are usually high enough to make NIR lines undetectable, then the search for such tracers has shifted toward even longer wavelengths. 

In the MIR regime, \citet{Rigliaco15} have found for PMS stars a correlation between the intensity of the Humphreys ${\, \alpha}$ line (denoted Hu${\, \alpha}$ or \hi 7$-$6) at 12.372 $\mu$m with the luminosity of the H${\alpha}$ line in the optical (calibrated in \citealt{Rigliaco12}) using Spitzer Space Telescope.
This demonstrates that other \hi lines can similarly be used as tracers of the accretion luminosity. Recently with JWST, it has been possible to establish new MIR \hi lines as \Lacc{} tracers (\citealt{Tofflemire2025} and \citealt{Shridharan2025}), adding a total of seven \hi lines (between the range 5.9 $-$ 19 \mum{}) to the set of available tracers which we can, in principle, use to study accretion rates in embedded objects (see Table \ref{tab:hi_lines}).  

For low-mass Class II objects, the NIR \hi tracers yield $\dot{M}_{\rm acc} \lesssim 10^{-8} \, \rm M_{\odot} \, \rm yr^{-1}$ \citep{Alcala2014}, while for the intermediate-mass regime they point to $\dot{M}_{\rm acc}$ between $\sim 10^{-8}$ to $ 10^{-4} \, \rm M_{\odot} \, \rm yr^{-1}$ \citep{Grant2022, Wichittanakom2020, Rogers2025}, although sometimes with large dispersion. 
For the high-mass side we can refer to theoretical works, which suggest higher \Macc{} up to $ \sim 10^{-4} \, \rm M_{\odot} \, \rm yr^{-1}$ set by their periodic accretion outbursts, which should progressively decrease as the star evolves to the main sequence phase \citep{Hartmann2008}. As the embedded phases are hard to probe, most of the observational constraints come from their corresponding outflow ejection rates, because they are thought to represent a fraction ($\sim$1\%) of the accretion rates \citep{Shepherd1996a, Shepherd1996b, Beuther2002b, Arce2007, Frank2014}.

\begin{figure*}
\centering
\includegraphics[width=0.99\linewidth]{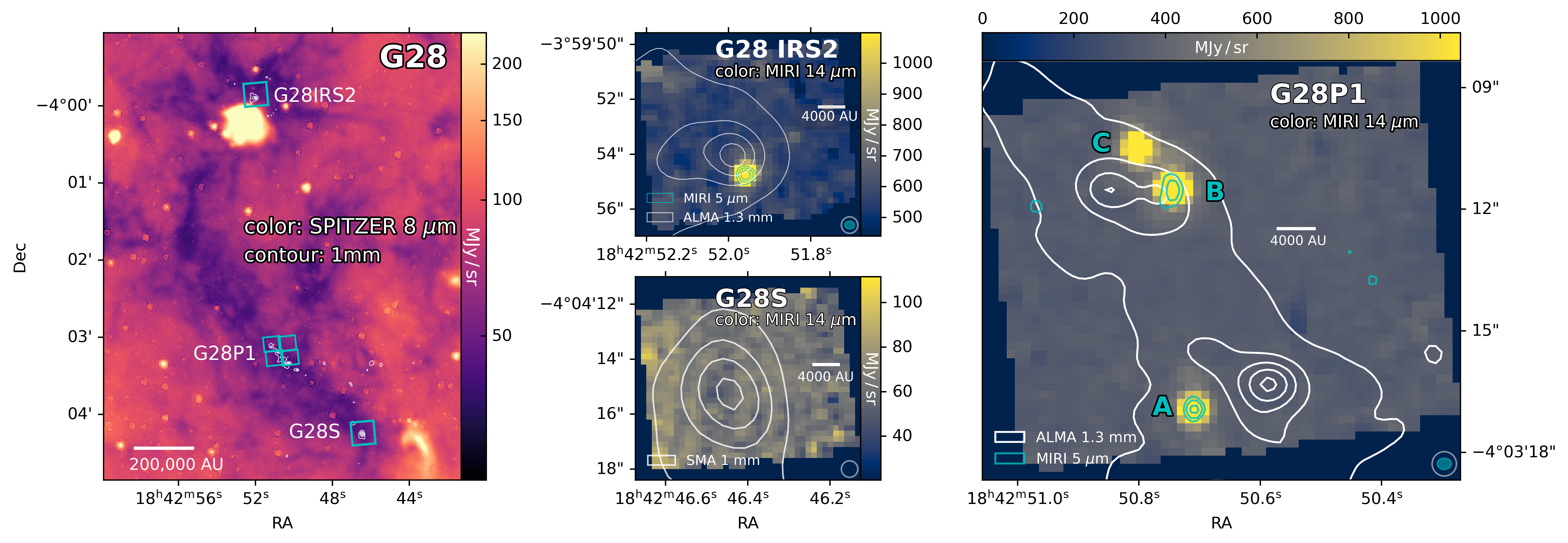}
\caption{Overview of the three less evolved regions of our sample: G28IRS2, G28P1, and G28S. In this case all of them belong to the same IRDC G28. Left: \textit{Spitzer} (8 $\mu$m) view of the whole IRDC G28. The three target subregions are depicted with millimeter contours (white) from ALMA for G28IRS2 \citep{Molinari2025} and G28P1 \citep{Zhang2015}, and from SMA for G28S \citep{Feng2016a}, with cyan boxes highlighting the field of view of our MIRI observations, although slightly magnified for better visualization. Middle and right panels: zoom-in to the target subregions. They show the MIRI continuum at 14 $\mu$m, whereas cyan contours trace the continuum at 5 $\mu$m (no IR source was detected toward G28S). The ellipses in the bottom right corners represent beam sizes,  grey for the 14 $\mu$m image, and cyan for the 5 $\mu$m contours.}
\label{Fig:Overview_G28}
\end{figure*}

\begin{figure*}
\centering
\includegraphics[width=0.99\linewidth]{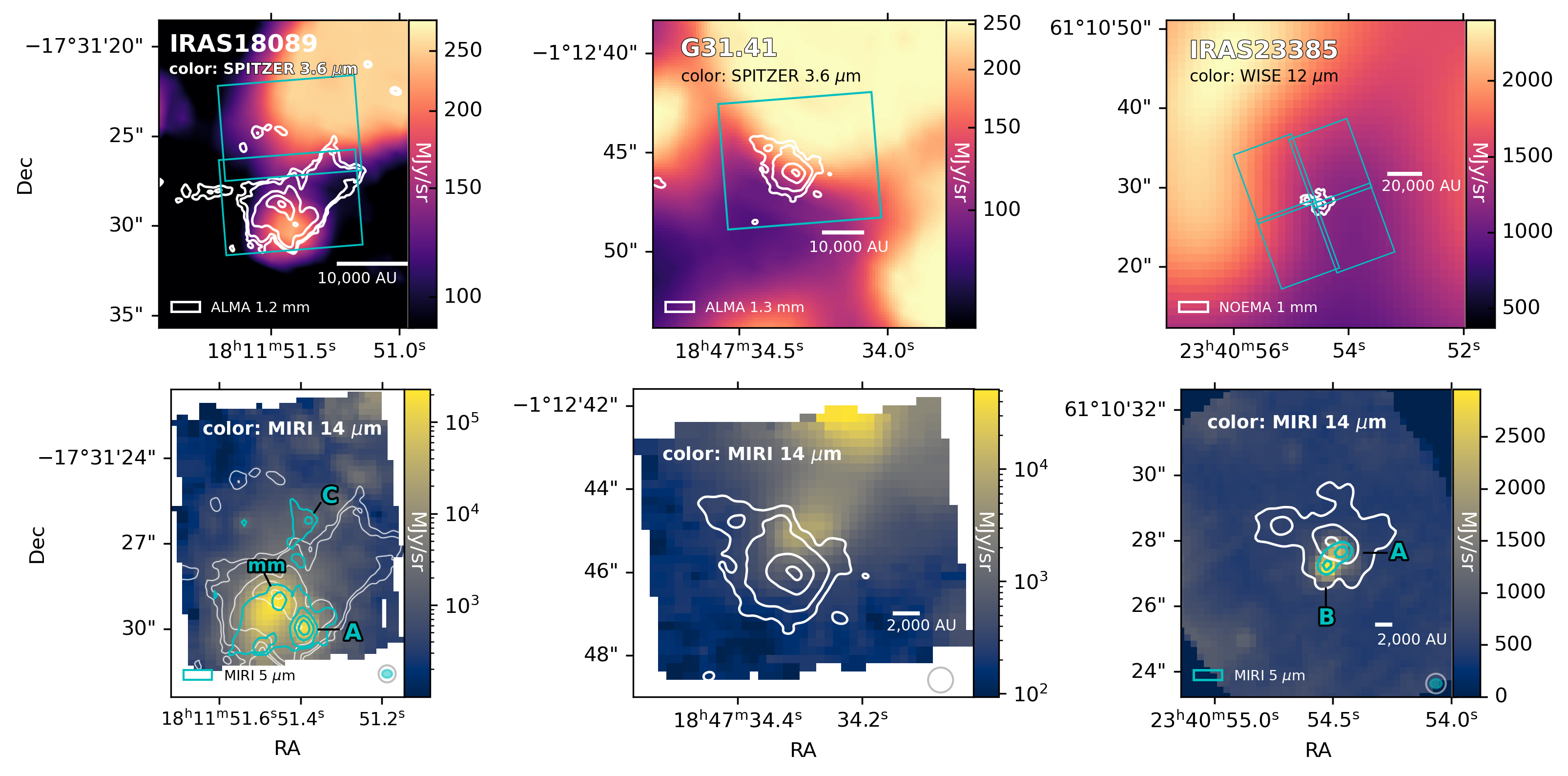}
\caption{Overview of the three more evolved protostellar regions of our sample: IRAS18089, G31, and IRAS23385. Top panels: they show each region as seen by \textit{Spitzer} (3.6 $\mu$m) or WISE (12 $\mu$m), with white contours depicting millimeter data from ALMA for IRAS18089 (1.2 mm; \citealt{Sanhueza2021}) and G31 (1.3 mm; \citealt{Beltran2018}), or NOEMA for IRAS23385 (1 mm; \citealt{Beuther2018}). The cyan boxes highlight the field of view of our MIRI observations. Bottom panels: MIRI integrated images at 14 $\mu$m for each region covering the corresponding upper panels. The cyan contours represent the continuum at 5 $\mu$m.}
\label{Fig:Overview_evolved}
\end{figure*}

\begin{table*}[t]
\caption{JOYS source sample from millimeter observations. }
\begin{tabular}{lrrlllllllr}
\hline
\hline
Archive name & RA$^a$ & Dec$^a$ & D & $L_{\rm bol}^{\,\,\,b}$ & Class & $M_{\rm env}^{\,\,\,b}$ & $v_{\rm lsr}$ & Mosaic  & Other names & Ref. \\
& [J2000] & [J2000] & (kpc) & (L$_\odot$)  & & (M$_\odot$)& (km$\,$s$^{-1}$) & & & \\
\hline
G28IRS2 & 18:42:51.99 & -3:59:54.0 & 4.51 & 10120    & IRDC/HMPO & 3288	&  77.7  & 1x1  & G28P2 & $^{3, 4, 12}$\\
G28P1 & 18:42:50.59 & -4:03:16.3 & 4.51 & 682       & IRDC & 4276 		& 80	& 2x2 &  - & $^{3,4, 10}$ \\
G28S & 18:42:46.45 & -4:04:15.2 & 4.51 & 364         & IRDC & 2296 		& 78.4  & 1x1  & C1-Sa & $^{3,4,6,7}$ \\
IRAS23385 & 23:40:54.49 & 61:10:27.4 & 4.90 & 3170   & HMPO & 220 		& -50.5 & 2x2 &  Mol160 & $^{1,2,11}$\\
IRAS18089 & 18:11:51.24 & -17:31:30.4 & 2.34 & 15723$^c$ & HMC &1100$^c$& 34.1  & 2x1  & G12.89+0.49 & $^{4,5,8,9}$\\
G31 & 18:47:34.33 & -1:12:45.5 & 5.16 & 69984    & HMC & 7889 			& 97	& 1x1  & - & $^{3,4}$ \\
\hline
\end{tabular}
\label{tab:mmsources}\\
References: 1:\cite{Beuther23}, 2: \cite{Molinari08}, 3:\cite{Wang08}, 4: \cite{Urquhart18}, 5: \cite{Xu11}, 6: \cite{Feng2016}, 7: \cite{Tan2016}, 8: \citet{Cesaroni1994b}, 9: \citet{Beuther2002a}, 10: \citet{Zhang2015}, 11: \citet{Cesaroni2019}, 12: \citet{Csengeri2016}. \\
$^a$ Coordinates of millimeter peak positions.  $^b$ Properties were derived for emission that extends over larger spatial scales than those of the IR sources we report in Table \ref{tab:IRsources} by factors of a few.   $^c$ Distance corrected.\\
\end{table*}

Given the newly available high sensitivity of the JWST, the Hu${\, \alpha}$ line becomes suitable for observations of embedded objects. 
In this line, \citet{Beuther23} reports a tentative 3-sigma detection of Hu$\, \alpha$ in the high-mass binary IRAS23385. Building on this initial study, we perform a more exhaustive search, where the goal is disentangling, for the first time, the accretion occurring at the very early stages of high-mass star formation. Our sample consists on six high-mass star forming environments, which can be classified as Infrared dark clouds (IRDCs), High-Mass Protostellar Objects (HMPO), and Hot Molecular Cores (HMC). 
Each of these regions are expected to host embedded stars with masses above $-$ or in the process of reaching $-$ 8 M$_{\odot}$.

To measure accretion rates, we study \hi accretion tracers available within the MIRI/JWST spectral coverage, assuming that the accretion luminosity relationships can be extended from Class II low-mass YSOs to high-mass protostars (but see discussion). We present the JWST data in Sect. \ref{Data}, and describe our methods in Sect. \ref{Section:Methods}. We show all the \hi lines that MIRI detects toward our regions, and present our results on accretion luminosities and rates in Sect. \ref{Section:Results}. We elaborate on relevant methodological uncertainties and limitations that can affect the accuracy of accretion luminosity (and rate) measurements in Sect. \ref{Discussion}, and we provide our conclusions in Sect. \ref{Conclusions}.


\begin{table}
\caption{IR compact sources at 5 $\mu $m detected in our MIRI observations.}
\begin{tabular}{llll}
\hline \hline
        \multirow{1}{*}{Archive Name} & \multirow{1}{*}{IR sources} & \multirow{1}{*}{RA} & \multirow{1}{*}{Dec (J2000)}  \   \\
        & & [J2000] & [J2000]\\
        		\midrule
        \multirow{1}{*}{G28IRS2} &  G28IRS2& 18:42:51.959 & -3:59:54.76  \\
        \midrule
        \multirow{3}{*}{G28P1}&   A &18:42:50.744 & -4:03:11.55 \\
         & B & 18:42:50.709 & -4:03:16.94 \\
         & C $^a$ & 18:42:50.801 &-4:03:10.53  \\
		        \midrule
        \multirow{1}{*}{G28S}& -& -& -   \\ 
        		\midrule
        \multirow{2}{*}{IRAS23385} & A& 23:40:54.468 & +61:10:27.65  \\
         & B & 23:40:54.522 & +61:10:27.26 \\
                \midrule
        \multirow{3}{*}{IRAS18089} & A &18:11:51.392  & -17:31:29.96 \\
         & mm & 18:11:51.452& -17:31:29.03 \\
         & C & 18:11:51.380 &-17:31:26.19  \\         
         		\hline 
        \multirow{1}{*}{G31} & - & - & - \\
        		\hline 
\end{tabular}
\label{tab:IRsources}\\
Notes: Source nomenclature is determined by brightness, starting from "A" for the brightest sources. 
Remarks: $^a$ Within FOV of channels 3 and 4 only. 
\end{table}

\section{Data} \label{Data}

This release is part of the MIRI GTO "JWST Observations of Young proto-Stars" (JOYS) program, which observed more than 20 star forming regions with the Medium Resolution Spectrometer (MRS), including six high-mass regions (PID 1290; PI: E. F. van Dishoeck). This work focuses specifically on their subsample of high-mass sources, which can be classified in two stages of evolution: the more active and hence evolved IRAS23385, G31, and IRAS18089, and the comparatively more quiescent, younger G28IRS2, G28P1, G28S. Figures  \ref{Fig:Overview_G28} and \ref{Fig:Overview_evolved} give an overview of our sample, where beam sizes are 0.3 and 0.6 arcseconds at 5 and 14~\mum{} respectively. The spectral resolving power ranges between 3500 to 3000 at 5 and 14~\mum{} respectively, and the typical noise level is $\sim 2\times10^{-5}$~Jy at the IR source locations.

\subsection*{Ancillary data} \label{section:Ancilliary-data} 

Previous observations at millimeter (mm) wavelengths (from 1 to 1.3 mm) provide general properties of our target regions (Table \ref{tab:mmsources} and references therein). In particular, we retrieved archival data from: ALMA for IRAS18089 \citep{Sanhueza2021}, G31 \citep{Beltran2018}, G28IRS2 \citep{Molinari2025}, and G28P1 \citep{Zhang2015}; NOEMA for IRAS23385 \citep{Beuther2018}; and SMA for G28S \citep{Feng2016a}. 
\section{Methods} \label{Section:Methods}

\subsection{Data reduction}
We run our JOYS- adapted version of the MIRI JWST pipeline (version 1.15.1), in line with all JOYS releases. Steps are detailed in \citet{van_Gelder2024a, van_Gelder2024b}, but as we ran an updated JWST pipeline version there are a few differences that we mention here. We included background subtraction on the detector level when available, namely for IRAS18089, and G31. In these cases, we masked science emission lines (i.e. H$_2$ 0-0$\,$S(1), H$_2$ 0-0$\,$S(2), [Ne$\textsc{ii}$]) on the background files prior to reduction. Despite it being also available for IRAS23385, we did not implement it because it contains significant emission of several species.
Although planned originally, background observations for G28IRS2, G28P1, and G28S were not taken. In all cases, the residual fringe correction and bad pixel correction tasks were turned on. The outlier pixel detection task was turned off. We note however that bad pixels are commonly seen, and represent an obstacle that cannot be fully overlooked. Some can be seen in Fig. \ref{Fig:map_spec1} (G28P1 A)  as a repeated pattern of pixels spread in the map. In our case, they are harmless as they appear out of our science apertures.
Lastly, for IRAS23385 in particular, we masked cosmic rays in the science files that appeared at the wavelengths of some Hydrogen emission lines.

\subsection{Astrometric correction} \label{method:astrometric_correction}
  
We used MIRI parallel images at 15.0 $\mu$m, their associated MIRI point source catalogues, and Gaia DR3 to identify and correct pointing deviations of JWST relative to Gaia. We found matching sources for MIRI to Gaia for three regions (G28IRS2, G28S, and G31), which revealed on average a small systematic shift of the JWST pointing of $\sim 124$ mas in right ascension, and 30 mas in declination. Such an average shift is, in fact, comparable to the angular resolution of JWST at 5 $\mu$m (130 mas). We corrected our MIRI astrometry using the individual values found for each source and, for the remaining three regions, we adopted different approaches that we describe in detail in Appendix  \ref{Appendix:Astrometric-correction}.

\subsection{Continuum subtraction} \label{Method:ContinuumSubtraction}

To study line features we performed a pixel-wise subtraction of the continuum around each line. To do so, we first masked a small spectral range at the science lines, and then obtained the best fit among polynomial functions of order 1, 2, and 3,
 using a root mean squared error (RMSE) criterion. The selected function is finally subtracted from the corresponding original pixel spectrum. 
The 2nd- and 3rd- order polynomials were strictly necessary for IRAS18089 (Appendix \ref{Appendix:contsubIRAS18089}), and also relevant for the bright G31 outflow, but we still include them for the other regions for consistency, and because they still offer slightly better continuum fits. 

As so far described, the procedure achieved clean line maps for five out of our six regions, but for IRAS18089 a few more considerations were necessary. We provide details for IRAS18089 in Appendix \ref{Appendix:contsubIRAS18089}. 
 Although the subtraction procedure could not be applied equally to all regions, we emphasize that we achieved data cubes with homogeneous spectral noise across the entire sample (up to 10$^{-4}$ Jy).

\subsection{Searching for accretion tracers: Strategy} \label{Method:strategy}
\hi emission lines contain key information about the accretion processes occurring in a protostar. We searched for all the  Hydrogen lines that can be detected given the sensitivity of our MIRI/MRS observations (from 5 to 28 $\mu$m). To achieve this, we tested the 25 \hi lines in that wavelength range that involve a maximum of 5 quantum level transitions of the electron, namely $\Delta n = n_{\rm u} - n_{\rm l} \leq 5$. We introduce them organized by increasing wavelength in Table \ref{tab:HI_wavelengths}. Transitions involving larger values of $\Delta n$ are most likely not-detectable given our sensitivity. To also account for line contaminants, we tested the presence of water line emission as they are often found toward Class II sources \citep{Rigliaco15}. 

We use a typical radius of $0.^{\prime \prime}5$ to set the protostar apertures when extracting their spectra, both for estimating extinctions (Sect. \ref{Method:extinction}) and to study emission lines (Sect. \ref{Result:detecting-HI-transitions}). Simultaneously, we evaluated their local background as an annulus centered on the protostar with a typical radius of $1.^{\prime \prime}5$ (excluding the protostar's aperture), although in some cases, to correctly represent the local ambient conditions, we arbitrarily shifted it off from the protostar position or varied their angular sizes (between $1.^{\prime \prime}2$ and $1.^{\prime \prime}8$). This is to avoid the noisier image edges (e.g. G28P1 in Fig. \ref{Fig:map_spec1}) or areas affected by strong ionization driven by external sources (e.g. IRAS23385 or IRAS18089 in Figures \ref{Fig:map_spec2} and \ref{Fig:map_spec4} respectively). 
We compute the signal-to-noise ratio (SNR) as the line emission peak over the baseline uncertainty, and additionally introduce a protostar-to-background line luminosity $L_{\rm proto} / L_{\rm bkg} $ ratio (Table \ref{tab:fitted_accretion_lines}) to properly assess the significance of compact detections toward the protostars. Then, we classify an \hi line as a "compact detection" only when SNR $\geq 3$ and $L_{\rm proto} /L_{\rm bkg}>1 $.

\subsection{Extinction A$_{\lambda}$} \label{Method:extinction}

We estimated extinction at a given wavelength $A_{\rm \lambda}$ in two steps: we first obtained $A_K$ for each protostar from the absorption depth of the silicate feature ($\tau_{9.7}$), and then converted A$_K$ to the wavelength of interest $A_{\rm \lambda}$ (\hi lines) using the two extinction curves of \citet{Pontoppidan2024} and \citet{McClure09}. We hereafter refer to them as the KP5 and M09 extinction curves, respectively.  Our specific approach follows that of \cite{Gieser2026subm}, which builds up upon the analysis of \cite{Rocha24}. For a summary of ways to measure extinction see \citet[Appendix G]{van_Dishoeck2025}.

\begin{figure*}[h]
\centering
\includegraphics[width=.91\textwidth]{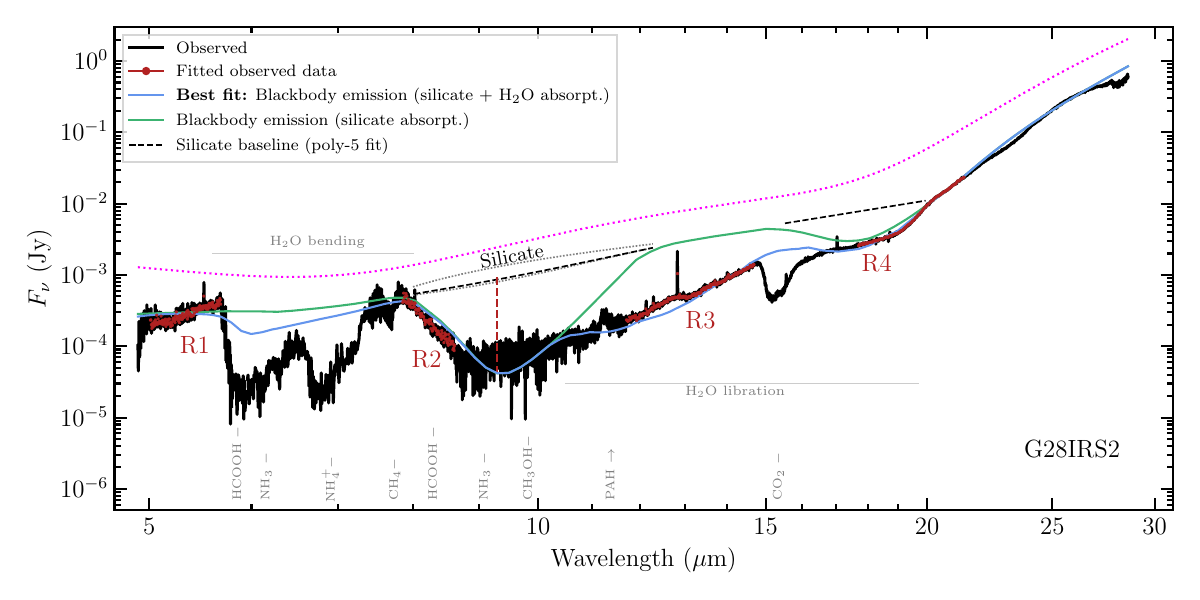}
\caption{Full MIRI spectrum of the G28IRS2 protostar (black). It shows several absorption features, where the broader one at 9.7 $\mu$m is produced by silicate grains. Blackbody components (3 in this case) absorbed by silicates and H$_2$O are fitted to reproduce the overall continuum. The blue curve represents the fit obtained by considering those spectral ranges that are predominantly absorbed by silicates and H$_2$O (in red, also indicated by the R1-R4 ranges). The green curve represents blackbody emission that would only be absorbed by silicates, while the dotted magenta curve shows the modeled blackbody emission unaffected by absorption. The black dashed curve above the main silicate absorption feature at 9.7 \mum{} is an interpolation of the silicate-absorbed continuum from a five-order polynomial fit. Here, the red vertical line highlights the resulting depth of the feature that determines $\tau_{9.7}$, whose uncertainty is given by other two polynomials (order 1 and 4) fits passing above and below (grey curves). The black dashed line at R4 indicates the location of the secondary silicate absorption feature.}
\label{Fig:silicate_fit_G28IRS2}
\end{figure*}

In step one, we used the same sky aperture used to study the protostellar spectra (Sect. \ref{Method:strategy}). The main absorption feature of silicates at 9.7 \mum{} is mainly produced by silicate grains along the line-of-sight towards the central source, and its depth is what we aim to measure. We modeled silicate dust compositions as mixtures of Olivine and Pyroxene following the Synthetic silicate models of \citet{Rocha24}. As the bending mode (secondary) feature of silicates at $\sim$ 18 $\mu$m is mixed with water ice absorption, we included water ice models at temperatures of 15 and 75 K following \citet{Gieser2026subm}. These two are the only absorbing species that we model. In parallel, we modeled the emitting source as that of two blackbody components representing warm emission due to the embedded protostar, and cold emission due to the dusty protostellar envelope. Only for G28IRS we required adding a third (hot) blackbody component. Lastly, we provided the fitter with spectra masked to include only the four spectral ranges indicated in Fig. \ref{Fig:silicate_fit_G28IRS2} (R1 - R4), because at those wavelengths the water and silicate absorption should dominate over that of other species. These input spectra are also rebinned by a factor of five to smooth out noisy ranges or strong emission lines.  More in detail, range R1 is defined where we expect minimal absorption features, thus it provides the closest look to the global IR-continuum. Range R2 represents the short-wavelength slope of the main silicate feature, expected to be dominated by silicates, extending toward the feature peak but before reaching the noise level. R3 is defined where H$_2$O (libration) absorption becomes dominant over the silicates up to the prominent CO$_2$ absorption feature at $\sim$15 \mum{}, and avoiding the broad PAH emission feature (at 11.28 \mum{}) when present. This also means that these ranges are adapted for each region independently. Lastly, R4 marks the secondary silicate feature that still has some H$_2$O libration absorption. 

Given this setup, the fitter used non-linear least squares to optimize a total of 8 parameters: two blackbody temperatures with their corresponding scaling factors that represent their emitting surface sizes (3 blackbodies for G28IRS2); abundances of Olivine and Pyroxene; and abundances of H$_2$O  at two temperatures (15 and 75 K; Table \ref{tab:extinction_fitting_parameters}). In Sect. \ref{Discussion:extinction} we discuss the structure and properties of these protostellar systems as derived from the models. For a better representation of the noisy ranges between 8.8 and 11 $\mu$m or between 5.8 and 7.4 $\mu$m, some sources' spectra were binned by factors of 4 (IRAS18089C), 5 (G28P1 A and B) or 10 (G28IRS2).

To obtain the depth of the main silicate feature at $\lambda \sim 9.7 \, \mu$m ($\tau_{9.7}$), we estimated its baseline through polynomial fits connecting the resulting silicate absorbed spectrum (green curve in e.g. Fig. \ref{Fig:silicate_fit_G28IRS2}) from 7 to 14 $\mu$m but masked at the silicate feature between 7.6 and 12.4 $\mu$m. Depending on the source, the selected polynomial has order 3, 4, or 5, and it is always represented as a black curve (Figures \ref{Fig:silicate_fit_G28IRS2}, \ref{Fig:sillicate_fits2}, and \ref{Fig:sillicate_fits3}). The depth of the silicate feature $\tau_{9.7}$ is illustrated with a vertical dashed line, and we calculated it using

\begin{equation}
\tau_{9.7} = -ln \left(  \frac{F_{9.7}^{\rm \,model}}{F_{9.7}^{\rm \,base}} \right)
\end{equation}

where $F_{9.7}^{\rm \, base}$ is the baseline flux passing above this absorption feature (as just described), and $F_{9.7}^{\rm \,model}$ is the flux at its minimum. We note here that, in several cases, this fitted minimum appears below the noise floor, which occurs because the noise floor is hiding the tip of the main silicate feature in all our sources. When the fit does not go deeper than the noise floor, one might consider its derived $\tau_{9.7}$ as a lower limit (possibly IRAS18089 A and C). To estimate the uncertainty on $\tau_{9.7}$, we adopted other two polynomial fits (order 2, 3, 4, or 5) that fit different baselines passing above and below the selected one, resulting in an asymmetric uncertainty for $\tau_{9.7}$. They are also displayed in Figures \ref{Fig:silicate_fit_G28IRS2}, \ref{Fig:sillicate_fits2}, and \ref{Fig:sillicate_fits3}.
We subsequently converted $\tau_{9.7}$ to $A_{\rm K}$ using the "dense cores" relation of \cite{Boogert13}, namely $\tau_{9.7}= (0.26 \pm 0.01)\times A_{\rm K}$. 

In step two, we converted A$_K$ to the corresponding wavelengths of the \hi lines we are interested in. For this, the two extinction models of KP5 and M09 can provide significantly different conversions depending on wavelength, and it is not clear which one suits best the environments we are studying in this work. Although M09 is proposed for $A_{\rm K} <7$ mag (below our estimations), we still applied both curves in further luminosity calculations (e.g. duplicated cells in Table \ref{tab:fitted_accretion_lines}) and discuss their implications. We note that the difference between two $A_{\lambda}$ values given by the two extinction curves drive a larger uncertainty on $A_{\rm line}$ than that of the individual measurements. For that reason, we use both values in Sect. \ref{Result:detecting-HI-transitions} to correct observed line luminosities $L_{\rm line, obs}$ using 

\begin{equation}
L_{\rm line} = L_{\rm line, \,obs} \cdot 10^{\, A_{\lambda} /2.5}
\end{equation}

where  $L_{\rm line}$ is the extinction- corrected luminosity. Both observed and corrected line luminosities are reported in Table \ref{tab:fitted_accretion_lines}. Similarly to $A_{\rm line}$, the uncertainty on $L_{\rm line}$ is given by the two values from the two extinction curves.

\section{Results} \label{Section:Results}

\subsection{Detected protostars} \label{Results:Detected-protostars}
Figures \ref{Fig:Overview_G28} and \ref{Fig:Overview_evolved} present our data and the larger scale environment they reside in as seen by Spitzer Space Telescope and WISE in the IR. They also display with contours the distribution of cold and dense material observed in the millimeter (Sect. \ref{section:Ancilliary-data}). Our data allows us now to identify precise IR source positions from the continuum peaks revealed by MIRI at 5 $\mu$m. As spatial resolution improves towards shorter wavelengths, the 5 $\mu$m contours are best to separate very nearby sources (e. g. the binary in IRAS23385, see Fig. \ref{Fig:Overview_evolved}, also in \citealt{Beuther23}). While Table \ref{tab:mmsources} lists the previously identified coordinates of these massive regions from millimeter observations, Table \ref{tab:IRsources} shows the coordinates of these newly detected IR sources. We note that there is full agreement between the 5 and 14 $\mu$m peaks (both shown in Fig. \ref{Fig:Overview_G28} and Fig. \ref{Fig:Overview_evolved}), but also toward longer wavelengths, indicating that scattering is not biasing the true positions of the IR sources.   

Within a single FOV we detect from zero to three IR continuum peaks, hence when they are two or three we add an extra label ('A', 'B', or 'C') in Table \ref{tab:IRsources} to mark them as identified IR sources, where we assign label 'A' to the brightest continuum source at 5 \mum{}. The overview Figures \ref{Fig:Overview_G28} and \ref{Fig:Overview_evolved} display the millimeter contours in all panels, allowing us to compare them with the IR emission at 14 $\mu$m (color scale) and 5 $\mu$m (cyan contours).  In most cases, the IR sources appear shifted from the millimeter peaks even after astrometric corrections (Sect. \ref{method:astrometric_correction}). We interpret this as evidence of distinct sources, where potential sources at the mm peaks are too embedded to be seen at IR wavelengths. In turn, this highlights the clustered nature of the regions and likely slightly different evolutionary stages within them. 

From now on throughout this work, we treat these detected IR sources as the protostar themselves, where we expect the accretion luminosity to come from.

\begin{table}
\caption{Extinction values from the $\tau_{9.7}$ depths.}
\begin{tabular}{lcc|cc}
\hline 
IR source & $\tau_{9.7}$  &$A_{\rm K}$ &  $A_{12.37}$ $^*$ & $A_{12.37}$ $^{**}$ \\ \hline  
\noalign{\vskip .1cm}
G28 IRS2 	& 3.1$^{+0.34}_{-0.07}$ &  11.9 $^{+1.38}_{-0.54}$ &  9.3 & 5.8 \\ \noalign{\vskip .1cm}
G28P1 A 	& 2.2$^{+0.034}_{-0.03}$ & 8.6 $^{+0.36}_{-0.35}$ & 6.8 & 4.2 \\ \noalign{\vskip .1cm}
G28P1 B 	& 3.9$^{+0.0003}_{-0.08}$ & 12.0 $^{+0.46}_{-0.54}$ & 9.4 & 5.9 \\ \noalign{\vskip .1cm}
IRAS18089 A & 7.9$^{+0.06}_{-0.12}$ & 30.3$^{+1.18}_{-1.2}$ & 23.8 & 14.9	 \\ \noalign{\vskip .1cm}
IRAS18089 C & 5.4$^{+0.03}_{-0.08}$ & 20.7$^{+0.80}_{-0.86}$ & 16.2 & 10.2	 \\ \noalign{\vskip .1cm}
IRAS23385 A & 2.1$^{+0.04}_{-0.001}$ & 8.1 $^{+0.34}_{-0.31}$ & 6.3 & 3.9 \\ \noalign{\vskip .1cm}
G28S 		& -   & - & -  & - \\ \noalign{\vskip .1cm}
G31 		& -   &  - &  - & - \\ \noalign{\vskip .1cm}
\hline
\end{tabular}\\
$^*$ converted from A$_K$ using \cite{Pontoppidan2024}.\\ 
$^{**}$ converted from A$_K$ using \cite{McClure09}.
\label{tab:extinction}
\end{table}

\subsection{Extinction A$_K$ from silicate absorption features} \label{Discussion:extinction}

Figures \ref{Fig:silicate_fit_G28IRS2}, \ref{Fig:sillicate_fits2}, and \ref{Fig:sillicate_fits3} show the full spectra of the protostars whose extinction is measured, together with the models of silicate- and water- absorbed emission that we fitted (Sect. \ref{Method:extinction}). Several other species that contribute to the absorption features are also indicated in the bottom of these figures. The final $\tau_{9.7}$ and $A_{\rm K}$ are reported in Table \ref{tab:extinction}, and the parameters of each fit are summarized in Table \ref{tab:extinction_fitting_parameters}. The errors reported for $\tau_{9.7}$ correspond to the uncertainty on the silicate baseline fit specifically.  Table \ref{tab:extinction} also shows that extinction models yield a factor of 1.6 difference at the wavelength of the \hi 7--6 line (12.37 $\mu$m), which reflects how important their disagreement can be. For G28S and G31 it was not possible to estimate $\tau_{9.7}$ because their sources are too embedded and the noise dominates their spectra.

One caveat on anchor range R2 (Sect. \ref{Method:extinction}) is that the continuum gets often affected by the nearby absorption of CH$_4$, HCOOH, other ices, or the redder (although already weak) tail of the broad H$_2$O ice bending absorption (for IRAS18089A see figure F.1. in \citealt{van_Dishoeck2025}; for detailed analyses of absorbing species at this narrow wavelength range see Chen et al., in prep.; and for a dedicated work on H$_2$O ice toward IRAS18089mm see \citealt{Gieser2026subm}).  In this regard, both G28P1 A and B appear to be the least affected by the latter species, and their fits seem to be the most reliable ones at R2.  In some cases R3 was misfitted (G28IRS2, IRAS18089 A and C, and possibly IRAS23385 A), which can be due to the noise floor hiding the lower tip of the absorption feature at 9.7 $\mu$m, together with the PAH emission, because they reduce the spectral range we can input to the fitter. Alternatively, water ice models at different temperatures can be tested, however, we consider that our fits achieve a significant precision that represent a step forward on measuring extinction toward the very dense protostellar envelopes.

On the dust composition side, the silicate features are mostly fitted with negligible quantities of Olivine compared to Pyroxene, where only for IRAS23385A they are comparable (Table \ref{tab:extinction_fitting_parameters}). Although more similar abundances of Olivine and Pyroxene can be expected,  other JOYS sources (low-mass, Class 0-I) are found to require similar ratios to fit the silicate feature at 9.8~$\mu$m (Chen et al., in prep.).

Although we are, in practice, interested in the extinctions, the obtained black-body (BB) parameters allow us to infer the properties of the material in the line of sight.
The two emitting components we modeled for each source consist of a warm black body component at temperatures between 349 and 678 K across regions, and a cold blackbody component at temperatures between 45 to 95 K (necessary to fit the secondary silicate absorption feature at $\lambda \sim 18 \, \mu$m, and most of the longer wavelength tail of the spectra). Only for G28IRS2 a third, hot component was required to fit range R1 (see Table \ref{tab:extinction_fitting_parameters}).  
From warm to cold, the components have progressively larger surface areas, suggesting that the warm blackbodies probe an innermost layer, while the cold blackbodies probe the external (cold) dust volume that possibly comprises most of the dust that drives the absorption observed in the warm (and hot for G28IRS) components.
All in all, this discrete approximation of what is, in reality, a continuum, is consistent with the notion of a system heated inside out.

\begin{figure*}

\includegraphics[width=.8\linewidth]{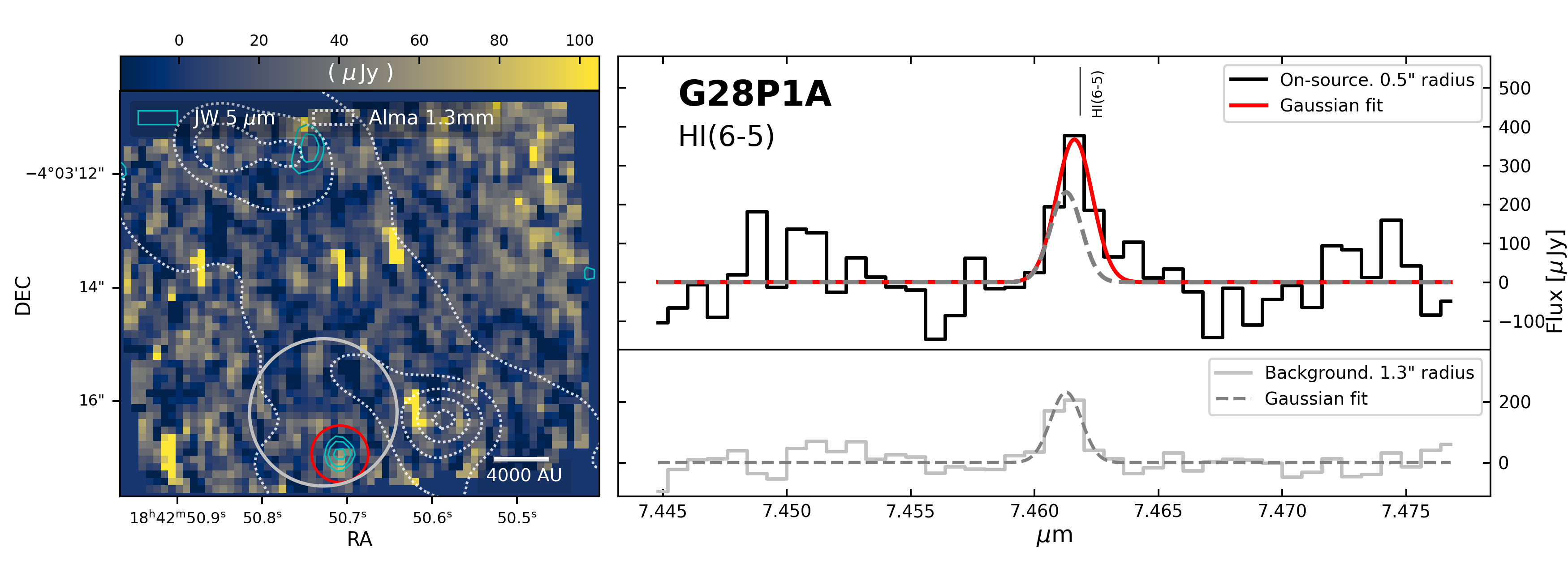}
\includegraphics[width=.8\linewidth]{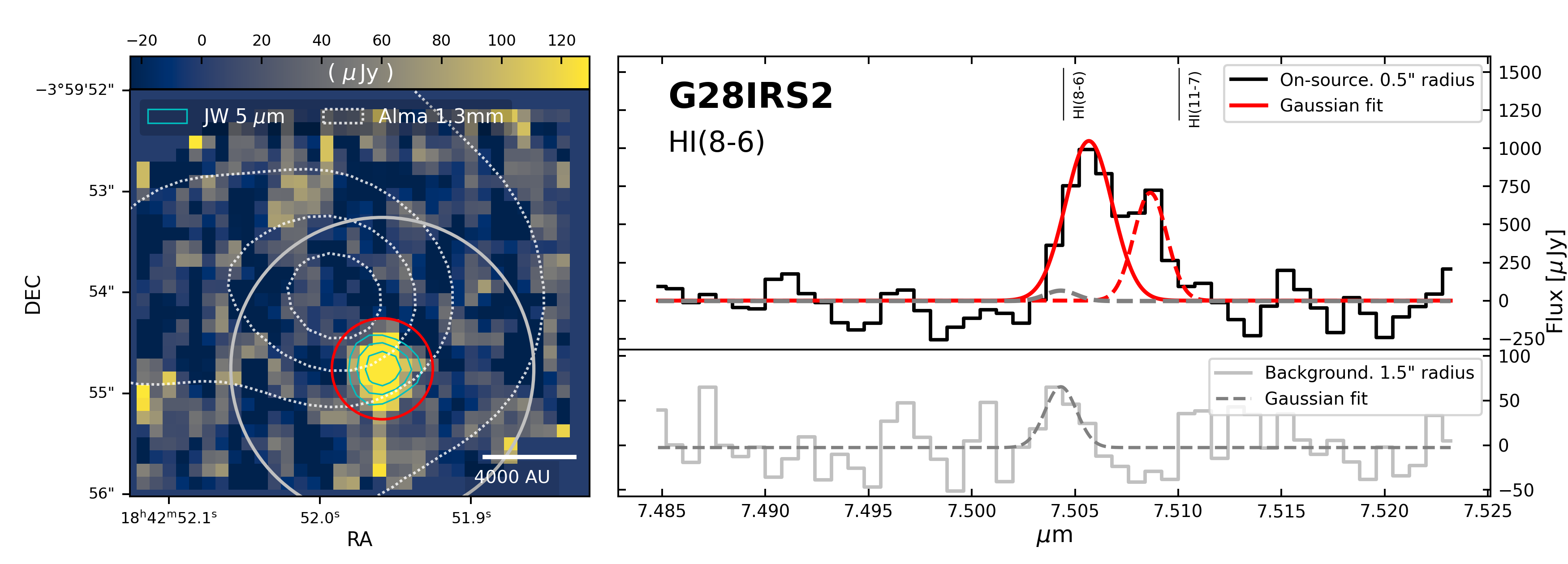}
\includegraphics[width=.8\linewidth]{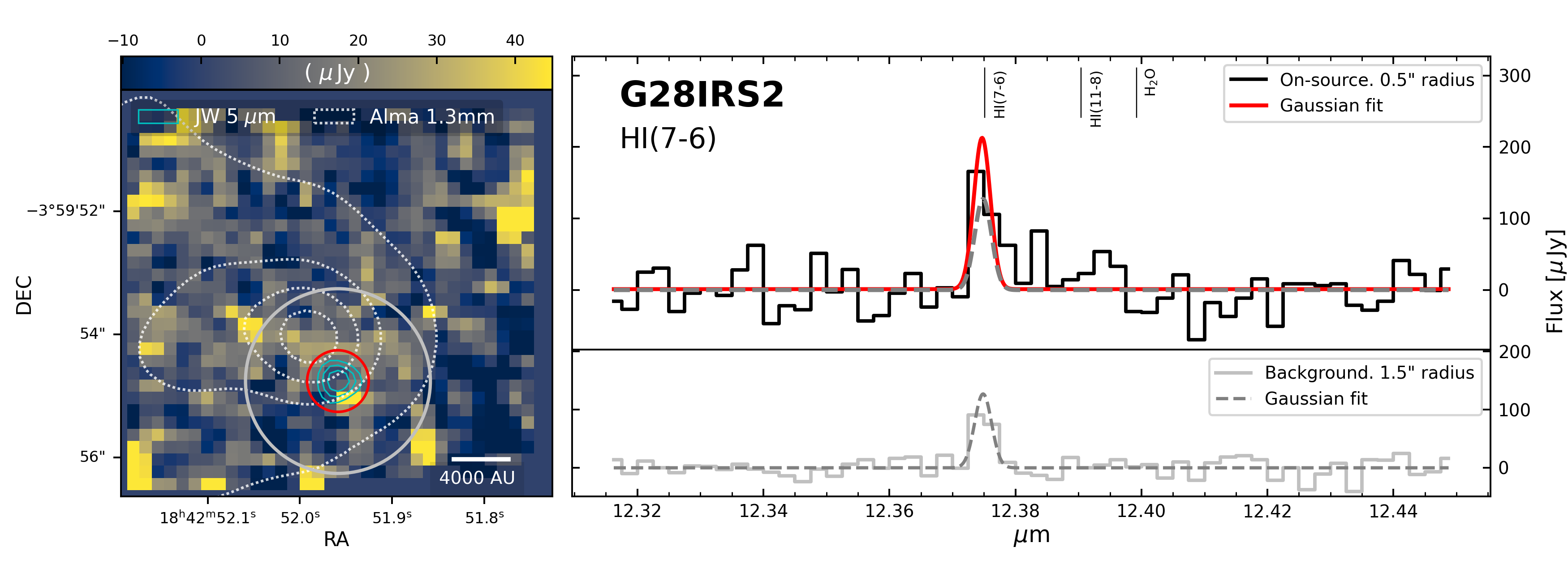}

\caption{The three most commonly detected \hi lines across our sample. We show them using two of our regions. Left: line moment 0 maps. The channels used to integrate the emission are those encompassed by the line Gaussian fits in the right panels. Red apertures enclose the map area used to extract the spectra at the protostars shown in right panels (upper insets), while the broader grey aperture (minus the central red aperture) is used to extract the spectra of the local background (right panels, lower insets). Right: Gaussian curves fit the line emission for both protostar (red, upper inset) and backgrounds (grey, lower insets). We reproduce the Gaussian fits of the background in upper insets to allow comparisons with the protostar Gaussian fit. For some of the lines (middle row) a second Gaussian is required for fitting another nearby emission line.
Other lines detected are shown in Figures \ref{Fig:map_spec2}, \ref{Fig:map_spec3} and \ref{Fig:map_spec4}.
}
\label{Fig:map_spec1}

\end{figure*}

\subsection{Detecting \hi emission lines} \label{Result:detecting-HI-transitions}

We detect several \hi lines, both as compact emission (Sect. \ref{Method:strategy}) and as diffuse emission in the environment around each protostar. The noise for our compact detections is homogeneous among different lines, with levels between 3 to 10$\times 10^{-5}$ Jy across all regions, and for the background apertures it is slightly lower with levels between 1 to 5$\times 10^{-5}$ Jy. The SNRs are reported in Table \ref{tab:fitted_accretion_lines}.  We count a total of 12 detections above a SNR of 3, and only four compact detections with SNR above 5, although that of IRAS18089mm likely corresponds to a hyper-compact (HC) \hii  region (Appendix \ref{Appendix:detailed-HI}). At the same time, we detect diffuse line emission ('E' in Table \ref{tab:hi_lines}) in the vicinity of our protostars in nine cases above SNR of 3. 

In general, this diffuse component is easy to find at high significance for large apertures in our sample, and its presence seems to depend on the environment around each region. In particular, G28IRS2 is in the vicinity of a very active clump that might contribute to ionize the gas (right southern in Fig. \ref{Fig:Overview_G28}). G28S also shows diffuse \hi emission, yet it lacks compact IR emission. In this case, a nearby \hii  region is present (seen western in Fig. \ref{Fig:Overview_G28}), whose gas might be reaching the LOS toward this region. Overall, this diffuse \hi component surrounding the cloud is common in regions of massive star formation.

Table \ref{tab:hi_lines} shows the detection status of eight \hi lines, where one tick mark indicates that the line is detected as compact emission at the protostar location with $3<$ SNR $<5$, and two tick marks that the line is detected with SNR $\geq5$. We note that including detections with SNR $\geq$ 3 becomes statistically relevant because there are several of them in our sample and, collectively, they demonstrate that \hi lines can be detected with MIRI/JWST. Table \ref{tab:hi_lines} also distinguishes whether the emission clearly comes (partially or totally) from a process different from accretion: Outflow ('O'), diffuse Environment ('E'), and previously identified Hyper-Compact-H$\,\textsc{ii}$ region ('$\dag$') are considered, as well as whether a line is seen in absorption ('A') with SNR $>3$.

\begin{table*}[h!]
\centering
\caption{Detection status of the \hi lines observed toward our newly-identified protostars.}
\begin{tabular}{lllllllll}
\hline \hline \rule{0pt}{2ex}  

 & \hi 9--6 & \hi 6--5 & \hi 8--6 & \hi 10--7 & \hi 13--8 & \hi 7--6 & \hi 8--7 & \hi 9--8\\
 $\lambda$ [$\, \mu$m$\,$]         & 5.908  & 7.460   & 7.502   & 8.760    & 9.392    & 12.372  & 19.062    & 27.803    \\
\hline
G28 IRS2     & \ding{55}  & \ding{55} E  & \textcolor{darkgreen}{\Checkmark}  \textcolor{darkgreen}{\Checkmark} & \ding{55}  & \ding{55}  & \textcolor{darkgreen}{\Checkmark} \textcolor{darkgreen}{\Checkmark} & \ding{55} E  & \ding{55} \\
G28P1 A      & \ding{55}  & \textcolor{darkyellow}{\Checkmark}   & \ding{55} & \ding{55} E  & \ding{55}  & \textcolor{darkyellow}{\Checkmark}  E  & \textcolor{darkyellow}{\Checkmark}  & \ding{55}  \\
G28P1 B      & \ding{55}  & \ding{55}    & \textcolor{darkyellow}{\Checkmark} & \textcolor{darkyellow}{\Checkmark}$^{!!}$  & \ding{55}  & \textcolor{darkyellow}{\Checkmark} E & \ding{55}   & \ding{55} \\
IRAS23385 A  & \ding{55}  & \textcolor{darkyellow}{\Checkmark}   & \textcolor{darkgreen}{\Checkmark}\textcolor{darkgreen}{\Checkmark} E  & \ding{55}   & \ding{55}  & \textcolor{darkyellow}{\Checkmark}$^{\star}$ & \ding{55} & \ding{55} \\
G28S         & \ding{55}  & E            & \ding{55} & \ding{55}   & \ding{55}  & E            & \ding{55}   & A \\
G31          & \ding{55}  & O           & \ding{55}& \ding{55}   & \ding{55}  & O            & O           & \ding{55} \\
IRAS18089 A  & - & -            & -         & \ding{55}   & -         & \ding{55}     & -           & - \\
IRAS18089 mm  & -  & \textcolor{darkgreen}{\Checkmark} \textcolor{darkgreen}{\Checkmark}$^{\dag}$ & - & \ding{55} & \ding{55}  & \ding{55} & A           & - \\
IRAS18089 C  & \ding{55}  & \ding{55}    & \ding{55} E & \textcolor{darkyellow}{\Checkmark}$^{!!}$ & \ding{55} & \ding{55}    & \ding{55}   & \ding{55}  \\
\hline
\hline \rule{0pt}{2.5ex}
$L_{\rm acc}$ cal. ref. & \cellcolor{white} $^3$ & \cellcolor{softgreen} $^3$ & \cellcolor{softgreen} $^3$  & \cellcolor{softgreen} $^{2,\, 3}$ & \cellcolor{white} $^3$ & \cellcolor{softgreen} $^{1, \,2, \,3}$ & \cellcolor{white} $^2$ & \cellcolor{white} $^-$ \\
\hline
\end{tabular}

\label{tab:hi_lines}

\rule{0pt}{3ex} 
Notes: The last row marks with green cells the lines whose calibration was used in this work to quantify accretion parameters.
Symbology:  \textcolor{darkyellow}{\Checkmark} line detected at protostar with $3<$ SNR $<5$ $-$ \textcolor{darkgreen}{\Checkmark}\textbf{\textcolor{darkgreen}{\Checkmark}} detected at protostar with SNR $\geq$ 5 $-$ 'E' detected diffuse at environment  $-$ \ding{55} not detected at protostar $-$ 'O' detected due to outflow $-$ 'A' in absorption with SNR $>3$ $-$ '$^{\star}$' SNR of 2.5 (lower than reported in \citealt{Beuther23}) $-$ $^{\dag}$ likely hyper-compact \hii  region $-$ '-' undetermined (see Sect. \ref{Appendix:contsubIRAS18089}) $-$ $^{!!}$ Dubious detection (Appendix \ref{Appendix:detailed-HI}) \\
$L_{\rm acc}$ calibration references. 1: \citet{Rigliaco15}, 2: \citet{Tofflemire2025}, 3: \citet{Shridharan2025}. 
\end{table*}

Among this set of lines, only five are detected in at least one region as a compact component with SNR $\geq 3$ at the location of the protostar: They are \hi 6$-5$, 8$-6$, 10$-7$, 7$-6$, and 8$-7$. Relevant caveats apply for the \hi 10--7 detections in particular (details for this and other detections in Appendix \ref{Appendix:detailed-HI}), hence we exclude those from the discussions in Sect. \ref{Discussion}. This is the emission one needs to quantify accretion luminosities (and rates later on), and most importantly, they all possess a \Lacc{}-calibration in literature. We include in Table \ref{tab:hi_lines} the \hi 13$-$8 and \hi 9$-7$ lines despite they show zero detections because they also have such calibrations in literature, and also included \hi 9$-$8  because it is generally a line of interest for being an alpha transition. 

For all of the lines detected as compact emission towards a protostar in Table \ref{tab:hi_lines} (one or two tick marks) we provide the corresponding continuum subtracted map and spectrum in Figures \ref{Fig:map_spec1}, \ref{Fig:map_spec2}, \ref{Fig:map_spec3} and \ref{Fig:map_spec4}. The left panels show the integrated line maps, including in red the apertures used to extract each protostar spectrum, and in grey the apertures for the local background. 
The right panels are divided into two insets to properly show both the protostar (upper insets) and background (lower insets) spectra, where a red curve shows our Gaussian fits for the corresponding \hi line. Since the background spectra are obtained from a three times larger flux-collecting area, we display them normalized to the protostar aperture size and reproduce them in the upper inset to enable a fair comparison. In some cases, nearby lines are necessary to fit in conjunction, hence we mark the wavelength location of all \hi (or H$_2$O) lines, corrected by their $v_{\rm lsr}$, on the upper part of each panel. Although the lower insets always show Gaussian fits to background features, we only mark them with the symbol 'E' in Table \ref{tab:hi_lines} if they surpass the SNR $>3 $ threshold. We note for completeness that the \hi 11--8 line is tentatively detected diffuse in IRAS23385 A, and possibly around G28P1 A (Fig. \ref{Fig:map_spec2}, third and fourth rows respectively).

\begin{table*}
\caption{Accretion parameters for the detected \hi lines derived using Gaussian fits (Figures \ref{Fig:map_spec1}, \ref{Fig:map_spec2}, \ref{Fig:map_spec3}, and \ref{Fig:map_spec4}).
}

\begin{tabular}{llllrr|ccc|ccc}
\hline \hline
\rule{0pt}{2.5ex}  
       \multirow{1}{*}{IR source} & SNR$^{\,\,*}$ & FWHM$^{**}$ & $\frac{L_{\rm proto}}{L_{\rm bkg}}$ & $L_{\rm line, \,obs}$ & $L_{\rm line}$ &  $L_{\rm acc}^{\,\,\,1}$  & $L_{\rm acc}^{\,\,\,2}$ & L$_{\rm acc}^{\,\,\,3}$ & $\dot{M}_{\rm acc}^{\,\,\,1}$  & $\dot{M}_{\rm acc}^{\,\,\,2}$ & $\dot{M}_{\rm acc}^{\,\,\,3}$   \\
         & &  [km s$^{-1}$]& & log[ $L_{\odot}$] & log[ $L_{\odot}$] & \multicolumn{3}{c}{ log[ $L_{\odot}$ ]} &    \multicolumn{3}{c}{ log[ $L_{\odot}$ yr$^{-1}$ ]}  \\ \midrule

\multicolumn{12}{c}{Pfund $\alpha$ \hi 6--5} \\ \hline         
\rule{0pt}{2.5ex}  
		G28P1 A & 4.5 (6.3) & 69 & 1.84 & -3.81  & -2.47 	&-&-&  2.02 &-&-& -5.2\\ 		 	
&		&  &                  &                  & -2.35 	&-&-&  2.16  &-&-& -5.1\\ \midrule  
        
		IRAS23385 A& 4.9 (7.2) & 164 & 15.7 & -2.92 & -1.66   	& -&- &  2.90  &-&-& -4.8 \\  
		&&&&										& -1.54 	& -&- &  3.02  &-&-& -4.7\\  

\midrule 

\multicolumn{12}{c}{Humphreys $\beta$ \hi 8--6} \\ \hline        
\rule{0pt}{2.0ex}  
		G28IRS2& 8.7 (-) & 107 & $>> 1$ & -2.83 & -0.96 & - & - &  3.84  &-&-& -3.4\\   
&		&  &                  &                 & -0.79 & - &-&  [4.03]   &-&-& -\\ \midrule  
        
		G28P1 B & 3.3 (-) & 108 & $>>$1 & -3.50  	& -1.62  & - & - & [3.12] &-&-&- \\   
		&&&                   &   					& -1.45 &   - & - & [3.31]  &-&-&- \\ \midrule 

		IRAS23385 A & 5.9 (4.4) &78 &2.27  & -3.44 	& -2.17 & - & - & 2.51 &-&-	& -5.2 \\   
		&&&                                & 		& -2.06 & - & - & 2.64  &-&-& -5.1 \\

\midrule
\multicolumn{12}{c}{Humphreys $\alpha$ \hi 7--6} \\ \hline
\rule{0pt}{2ex}  
		G28IRS2 & 7.4 (8.2) 	& 75 	&1.6  		&-4.30 	& -0.57 	&  [8.56] 	& [8.41] 	& [4.44] &-&-&- \\ 
		& &&                 						&   	& -1.96 	&   [5.66] 	& [5.62] 	& 2.95  &-&-& -4.3\\ 
 
        \midrule
		G28P1 A & 3.3 (9.2)& 72 & 1.17 & -5.00 & -2.30	& [4.96]	&   [4.96] 	& 2.59  &-&-& -4.6\\  
								&	&  &   &   & -3.31  & [2.86]   	&   [2.94]  & 1.52  &-&-& -5.7\\ \midrule   
	
        G28P1 B & 4.7 (15)& 70& 1.01 & -5.85 & -2.09 &  [5.40]& [5.37]& 2.82   	&-	   &-	  & -4.4\\    
        &		&  &         &               & -3.50 &   2.47 &  2.56 & 1.31 	& -5.1 & -5.0 & -5.9 \\ 
		        \midrule
		IRAS23385 A& 2.5 (0)& 90& $>>1$ & -4.24 & -1.70  & [6.21]  & [6.16]  & 3.24  &-  & - & -4.5 \\   
		&&&                   &                &  -2.65 &  [4.24]  & [4.26]  & 2.22  &    - & - & -5.5 \\  

\midrule 
\multicolumn{12}{c}{\hi 10 - 7 } \\ \hline      
\rule{0pt}{2.5ex}  
		G28P1 B$^{!!}$ & 4.2 (-) &177 & $>> 1$& -3.40 	& 0.47	&- & [7.20] & [5.22]  &-&-&- \\ 
        &		         &    &       &  		& -0.59	&-&  [5.78] & [4.16]  &-&-&- \\ \midrule 
	
		IRAS18089 C$^{!!}$ & 3.2 (-) & 123 &6.88  & -4.29  &  2.38  & -& [9.78] & [7.13]  &-&-&- \\ 
         &		             &     &      &        &  0.56  & -& [7.33] & [5.31]  &-&-&- \\ \midrule

\end{tabular}

\label{tab:fitted_accretion_lines}
\rule{0pt}{2ex}  
Notes:  Duplicated cells in last seven columns display the two values given by the extinction curves of KP5 (upper) and M09 (lower). The $L_{\rm acc}$ values in square brackets indicate that they surpass their region's $L_{\rm bol}$ (Table \ref{tab:mmsources}). \\Remarks:
$^{*}$ SNR toward the protostar and background (in parenthesis). $^{**}$ Not corrected by instrumental broadening. $^{!!}$ Dubious detection (Appendix \ref{Appendix:detailed-HI}).
\\$^1$ calculated using calibration of \cite{Rigliaco15}; $^2$ using calibration of \cite{Tofflemire2025}; $^3$ using calibration of \citet{Shridharan2025}.
\end{table*}

A particular obscured case is the G28S protostellar source because it is undetected at all MIRI wavelengths, despite its association with a compact outflow (see Sect. \ref{Discussion:outflows}).
Conversely, G31 shows a bright outflow lobe in the continuum  (Fig. \ref{Fig:Overview_evolved}, middle panel) and some \hi emission lines (e.g. Fig. \ref{Fig:G31_outflow_mosaic}), which makes it hard to study compact, accretion-driven emission on source (Sect. \ref{Discussion:outflows}).
Lastly, IRAS18089 is the most difficult case to interpret as detailed in Appendix \ref{Appendix:contsubIRAS18089}. We do not report any detections towards the brightest protostar IRAS18089 A, however we cannot yet rule out that this emission actually exists.

\subsection{Accretion parameters and luminosity from \hi lines}

Under the assumption that these compact detections are produced exclusively by accretion processes,
 one can convert their line luminosities $L_{\rm line}$ to accretion luminosities $L_{\rm acc}$ (we discuss potential contaminants in Sect. \ref{Discussion:outflows}). This is the prior step to estimate accretion rates, but also the less trivial one because it requires appropriate empirical \Lacc{}-calibrations. 
 All lines included in Table \ref{tab:hi_lines}, except \hi 9--8, have an available \Lacc{}- calibration. Still, we also exclude \hi 8--7 from further calculations because of the large uncertainty of its available calibration, especially at the higher \Lline{} regimes we obtain. Although these calibrations have been established for low-mass, Class II objects, we still apply them to our sample of embedded, high-mass sources, and we discuss their validity with the $L_{\rm acc}$ results at hand.

We measure the luminosities of the four Hydrogen lines of Table \ref{tab:hi_lines} that are detected to be compact, and Table \ref{tab:fitted_accretion_lines} shows their Gaussian fitting parameters. The protostar SNR value is accompanied with that of the corresponding background in parenthesis. When $L_{\rm proto}$/$L_{\rm bkg} >1$  we use the luminosity difference to report line luminosities ($ L_{\rm line, \,obs}$  $ = L_{\rm proto} - L_{\rm bkg} $) in Table \ref{tab:fitted_accretion_lines}. 
Next, we provide the extinction corrected $L_{\rm line}$, and the derived $L_{\rm acc}$, where two dividing vertical lines highlight that we are reporting all the possible $L_{\rm acc}$ values given the existing calibrations (when applicable). For all the extinction corrected quantities ($L_{\rm line}$, $L_{\rm acc}$, and $\dot{M}_{\rm acc}$) in Table \ref{tab:fitted_accretion_lines}, we provide a second value immediately below, where the upper one is derived using the KP5 extinction curve, and the lower one is derived using the M09 curve. We display $L_{\rm acc}$ in squared brackets to indicate when they surpass their region's $L_{\rm bol}$ (Table \ref{tab:mmsources}), and when they do not surpass $L_{\rm bol}$ we report their corresponding accretion rates in the last three columns. %

The \hi line widths (FWHM in Table \ref{tab:fitted_accretion_lines}) range from velocities of 69 to 177 km$\,$s$^{-1}$, where the MIRI spectral resolution corresponds 100 km$\,$s$^{-1}$, down to 80 km$\,$s$^{-1}$ \citep{Labiano2021}, thus some lines appear narrower than this resolution. There is no apparent correlation between specific lines and their widths. For instance, the compact detections of \hi 6--5 toward G28P1 A and IRAS23385 A show contrasting FWHM of 69 (effectively an upper limit of 80 km$\,$s$^{-1}$) and 164 km$\,$s$^{-1}$, which could be indicative of different formation mechanisms. 
While the magnetospheric accretion scenario points to velocity widths from 150~km$\,$s$^{-1}$ or more in Class II objects
(Tofflemire et al. 2025) and IRAS23385~A is consistent with that threshold, it is still unclear what the correct velocity widths should be toward massive protostars as the exact physics of the accretion process might be different than that of low-mass, Class II objects. %

\subsection{H$_2$O and OH lines}
Water commonly shows several emission lines in the MIR spectra of Class II disks (e.g. \citealt{Pontoppidan2010}), where some appear at the wavelengths of \hi emission lines like \hi 6--5, \hi~7--6, and \hi 12--7. For instance, \citet{Shridharan2025} found high median water contamination for these lines in a sample of 79 Class II disks, meaning that the \hi fluxes can get artificially enhanced if water emission is not taken into account. We therefore follow both \cite{Rigliaco15} and \cite{Tofflemire2025} to determine whether water is contaminating any of our protostars by observing relatively bright, isolated lines at 11.648, 12.396, 14.428, 15.17, and 17.22 $\mu$m, where the one at 12.396 $\mu$m is very close to the \hi 7--6 line.  

We found no emission of these water lines with SNR $\geq 3$ in our regions except by a single line at 14.428 $\mu$m in G28P1B, with a SNR of 5.8. However, among the lines we report for this particular region, the only detected line that can be affected by water is \hi~7--6, for which the nearby water line at 12.396 \mum{} is not seen (Fig. \ref{Fig:map_spec3}). This makes this tentative  detection of \hi~7--6 unlikely to be affected by water. Apart from this particular case, the evidence shows for the other regions that water is not contaminating our reported \hi line fluxes. We also note that the JOYS sample of low-mass, embedded sources (Class 0-I) show weaker lines \citep{van_Gelder2024b} than in the Class II sources of \citet{Rigliaco15} and \citet{Shridharan2025}.

We additionally searched for emission lines from the OH molecule, although with a different motivation. OH can potentially trace accretion because it might generate in the interaction between the UV radiation produced at the shock and the water budget of the inner disk \citep{Tabone2021}. OH line emission has been recently detected tracing accretion \citep{Watson2025}, while in previous studies it was detected from outflows (\citealt{Le_Gouellec2025}; Tychoniec et al. in prep.). Although there is currently no definitive method to quantify accretion luminosities or rates from OH, we tested whether OH is at least observed in our sample in favor of this outstanding debate. We searched for two quartets of OH lines at $\sim 16.0$ and 16.8 $\mu$m and found them only in IRAS23385 A (as already reported in \citealt{Francis2024}). They appear spatially compact at the protostar, with low SNR as shown in Fig. \ref{Fig:OH_IRAS23385A}. However, the fact that the four lines of each quartet are seen simultaneously provides certainty that the OH is present at the IRAS23385 A protostar. %

\subsection{Accretion rates} \label{Discussion:Accretion-rates}

Our reported accretion rates in Table \ref{tab:fitted_accretion_lines} are only tentative because they are derived from \Lacc{} estimations that already include important caveats (see discussion in Sect. \ref{Discussion:Caveats}). With this in mind, we offer the general picture.

Given a $L_{\rm acc}$ value, accretion rates $\dot{M}_{\rm acc}$ can be computed following Eq. (11.5) in \citet{Stahler_Palla2004} and solving it for \Macc{},
\begin{equation}
    \dot{M}_{\rm acc}  =   \frac{R_{\star} \, L_{\rm acc}}{G\,M_{\star} }
\end{equation}

for which the protostellar embryo mass $M_{\star}$ and radius $R_{\star}$ must be known. Within our sample, these stellar parameters are only known for IRAS23385 ($M_{\star}= 9$ \Msun{}, $R_{\star}= 5$ R$_{\odot}$; \citealt{Cesaroni2019}). For the rest of sources we can only use assumptions because existing (millimeter) observations for each region provide only the large scale core (or clump) masses which, in addition, tend to locate off from the actual IR core positions by at least a few arcseconds. We proceeded by adopting $M_{\star}= 5$ \Msun{} for all other sources, which is meant to represent sources that are still growing in mass and are bound to reach the 8 \Msun{} or more. Subsequently, we adopted  $R_{\star}= 9$ R$_{\odot}$ given the disk accretion model of \citet[Figure 16]{Hosokawa2010} for the given stellar mass, and assuming an accretion rate of order $\sim 10^{-5} \rm \, M_{\odot} \,$yr$^{-1}$.   

With the \citet{Shridharan2025} calibration providing most of our reported accretion luminosities in Table \ref{tab:fitted_accretion_lines}, tentative accretion rates range between $\sim 10^{-6}$ to $10^{-3}$ M$_{\odot} \,$yr$^{-1}$ depending on the region, tracer, and dust extinction model.

\section{Discussion} \label{Discussion}

\begin{figure}
\centering
\includegraphics[width=1\linewidth]{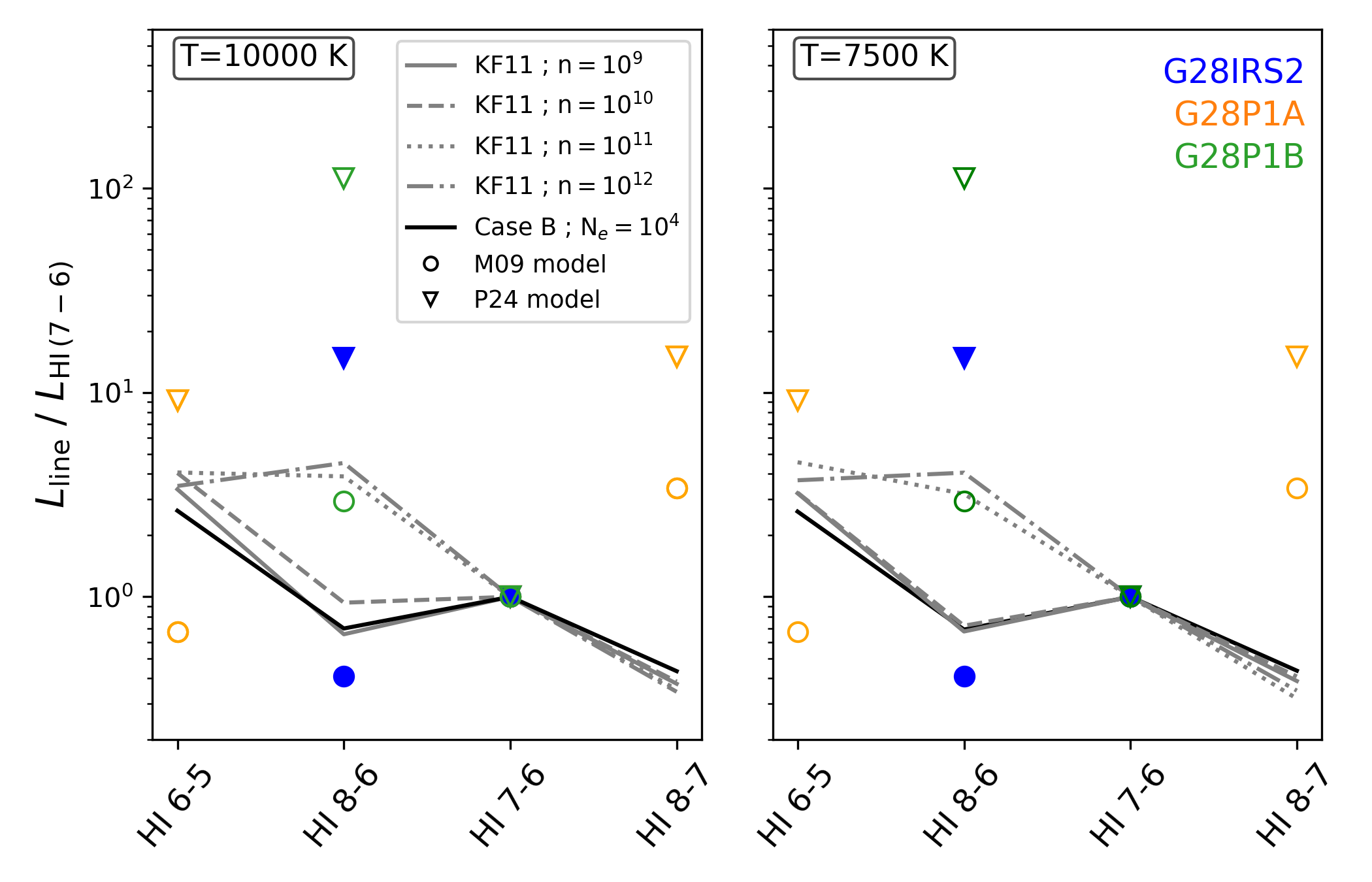}
\caption{Line ratios normalized to the Hu$\alpha$ line, compared to the local excitation (KF11) and Case B models \citep{Storey-Hummer1995}. Error bars are omitted, since the overall uncertainty is dominated by the discrepancy between the two extinction curves (triangle and circle markers). Solid markers represent detections with SNR $>5$, while open markers are detections with $3<$ SNR $<5$. Panels display models for temperatures of 10000 K (left) and 7500 K (right), and for the KF11 model we also include different density values ranging from n$_H = 10^9$ to $10^{12}$ cm$^{-3}$. We show the Case B model for a single electron density value $N_{\rm e}$ (in cm$^{-3}$) as this parameter appears to not change predictions significantly. Lines in the x-axis are organized by increasing wavelength.}
\label{Fig:line-ratios}
\end{figure}

\subsection{\hi line ratios compared to \hi excitation models} \label{Subsec:accretion?}

We compare  in Fig. \ref{Fig:line-ratios} the luminosity ratios of the observed lines normalized to the Hu$\,\alpha$ line (when detected with SNR $>3$), against Case B and CE models (see Sect. \ref{Intro}). We include two temperatures of 7500 and 10$^4$ K in separate panels in Fig. \ref{Fig:line-ratios}, and for the CE model in particular we include several densities. We find large dispersion in the measured line ratios, for which the models can hardly constrain particular conditions of temperature and density.  Although few cases may be well fitted by the models (mainly G28P1B), it still depends on the adopted dust extinction curve. For instance, the G28IRS2 protostar has both \hi 7--6 and \hi 8--6 lines detected with good SNR (solid markers), but its \Lline{} uncertainty given by the two extinction models of M09 and KP5 (circle and triangle, respectively) point to densities that are separated by orders of magnitude, namely higher than $n_H = 10^{12}$~cm$^{-3}$ (using KP5), and lower than $n_H = 10^{9}$~cm$^{-3}$ (using M09).

Case B and CE models displayed in Fig. \ref{Fig:line-ratios} also help to assess qualitative aspects of our detections in Table \ref{tab:hi_lines}. They show that the \hi 6--5 emissivity is expected higher than that of other $\alpha$-transitions (i.e. \hi 7--6 and 8--7), regardless of the temperature (Case B and CE) or density (CE). On top of this, extinction curves indicate that \hi 6--5 is the line least affected by extinction among those we detect here, hence it should be enhanced compared to, for instance, \hi 7--6.
Despite these expectations, the luminosities $L_{\rm HI\,6-5}$ are not among the highest, and the \hi 7--6 line is found at least in the same number of cases. 
Following the same idea, for the two regions where both \hi 6--5 and \hi 7--6 lines are detected, we find the expected $L_{\rm HI\,6-5} >$ $L_{\rm HI\, 7-6}$ trend for three out of four measurements (two ratios per extinction curve), but conversely the two \hi 8--7 measurements for G28P1A are far above any model predictions. 

Another interesting comparison involves both \hi 6--5 and \hi~8--6 as they are affected by very similar magnitudes of extinction due to their proximity in wavelength. Despite this mitigates the role of the adopted extinction model, their detections do not correlate well, because among the four regions where we detect at least one of them, the other is not detected in three cases. Part of the reason could be attributed to the low SNR of our detections, but on the other hand, the CE model predict inversions of the \hi 6--5 to \hi 8--6 flux ratio depending on density (Fig.~\ref{Fig:line-ratios}). While for low densities (log(n$_H$) $\lesssim$ 10) they predict fluxes $F_{\mathrm{H\,I}\,6\!-\!5} / F_{\mathrm{H\,I}\,8\!-\!6} > 1$, for high densities (log($n_{\rm H}$) $\gtrapprox$ 12) they predict $F_{\mathrm{H\,I}\,6\!-\!5} / F_{\mathrm{H\,I}\,8\!-\!6} < 1$. This observed dispersion in line ratios involving \hi 8--6 is therefore expected in the CE model, and it is a plausible scenario in which the volume density differences of our regions will be of a few orders of magnitude. More generally, the inconsistencies identified here (assuming that line emissivity models are correct) can be attributed to the low SNR of our observations, thus more and deeper observations will be necessary to correctly assess these aspects.

\subsection{Can MIR \hi lines trace accretion in protostars?} \label{Discussion:validity-of-tracers}

Our sample provides evidence that it is now possible to detect \hi emission produced at the massive protostellar systems, shifting the important questions one step further: What is actually producing this emission?. Despite the accretion process being the most likely origin, there is still room for having compact emission from the outflow (Sect. \ref{Discussion:outflows}), winds, or an (HC-)\hii region.  %

\begin{figure}[t]
\centering
\includegraphics[width=\linewidth]{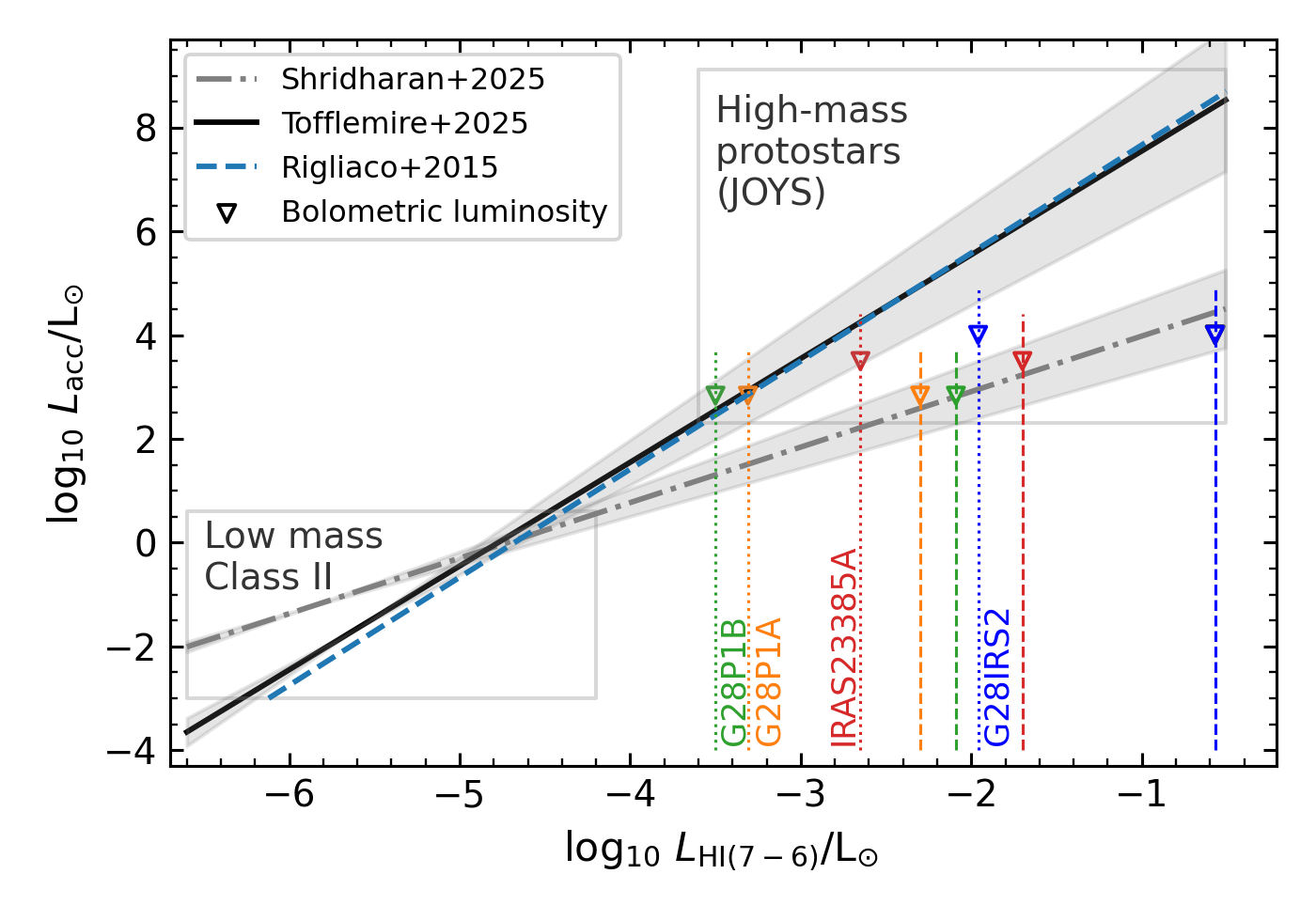}
\caption{Comparison of predicted $L_{\rm acc}$ (\textit{y-}axis) given three calibrations established for the Hu $\alpha$ line and our measured line luminosities $L_{\rm H \textsc{i} \,7-6}$ (\textit{x-}axis). The three curves reproduce these calibrations; the lower left box highlights the typical dynamical range over which they were derived, and the upper right box shows the contrasting line luminosity regimes of our sample, to which each calibration curve is extrapolated. The shaded areas correspond to the uncertainties of the \cite{Tofflemire2025} and \cite{Shridharan2025} calibrations. The colored vertical lines show our measured $L_{\rm H\textsc{i} \, 7-6}$ values (Table \ref{tab:fitted_accretion_lines}), where two per region are displayed given the two M09 (dotted) and KP5 (dashed) extinction curves. \Lbol{} values are indicated with triangles for each region, and corresponding $L_{\rm H\textsc{i} \, 7-6}$ are displayed up to 8 times each region's \Lbol{} to give a sense of the maximum \Lacc{} we can expect to still be physical (see Sect.~\ref{Discussion:validity-of-tracers}). Error bars are omitted, since the overall uncertainty is dominated by the two extinction curves 
}
\label{Fig:accretion_luminosity_HI76_summary}
\end{figure}

Based on their detection rate, the most suited lines to study accretion in massive protostars seem to be \hi 7--6, \hi 6--5, and \hi 8--6. However, the disagreement between available \Lacc{}-calibrations currently limits their use. To illustrate this point, Fig.~\ref{Fig:accretion_luminosity_HI76_summary} shows, for the \hi 7--6 line in particular, the distinct possible $L_{\rm acc}$ values given the three calibrations, and given both the KP5 and M09 extinction curves. Both \cite{Tofflemire2025} and \cite{Rigliaco15} (solid black and blue lines) calibrations return unphysically high accretion luminosities, as they go above their respective \Lbol{} (Table \ref{tab:mmsources}) in most cases. A more realistic picture is obtained with the \citet{Shridharan2025} calibrations, with many $L_{\rm acc}$ values below $L_{\rm bol}$ (14 cases out of 22, ninth column in Table \ref{tab:fitted_accretion_lines}), and some $L_{\rm acc}$ values above $L_{\rm bol}$ by factors of a few. In general, Figure \ref{Fig:accretion_luminosity_HI76_summary} well represents the cases of the \hi~6--5 and \hi 8--6 lines. %

We note two aspects on existing \Lbol{} measurements. (a) It is observationally possible that \Lbol{} is lower than the total luminosity by factors of a few. The total- to -bolometric luminosity ratio $L_{\rm tot}/L_{\rm bol}$ in a sample of Class 0/I/II objects increases with higher inclination angles and foreground extinction \citep{Furlan2016}. While inclination angles give median ratios from 1.5 to 3.5 (from low to high angles), $L_{\rm tot}/L_{\rm bol}$ reach up to 8.2 for sources with extinction $A_V > 50$ mag, as it is for most of our sources (we use this factor of $8\times L_{\rm bol} $ as an upper limit in Fig. \ref{Fig:accretion_luminosity_HI76_summary}). Along the same line, \citet{Pokhrel2023} found similar ratios due to increasing inclination ($\sim 1$ to 10), and  due to increasing extinction (2.1 to 6.1) for the Aquila and Orion clouds. In this scenario, \Lacc{} might account for most of the protostar's total luminosity by factors of a few, but it cannot explain the orders of magnitude differences that we often get given certain extinction model and \Lacc{}-calibration combinations. 
(b) These \Lbol{} measurements are obtained from larger spatial scales than those we use for our IR sources, which suggests that \Lbol{} may be overestimated for our protostars (\Lbol{} reported in \citealt{Molinari08} for IRAS23385; and in \citealt{Urquhart18} for the rest of our regions). Overall, these uncertainties on \Lbol{} do not allow our data to confidently disprove any specific \Lacc{}-calibration, but the higher \Lline{} cases strongly suggest that the relations of \citet{Rigliaco15} and \citet{Tofflemire2025} do not apply at the regimes of the higher-mass sources.

The works presenting the luminosity calibrations followed different approaches to establish their relations. While \citet{Rigliaco15} and \citet{Shridharan2025} obtain averaged curves from large samples, the \citet{Tofflemire2025} one studies a known binary system, benefiting from its predictable accretion burst periods. Although this closely guarantees a precise characterization of a single system, the work of \citet{Shridharan2025} counts with good self consistency between their different \hi line relations. On the other hand, recent works have found that the mechanism driving accretion might change from Herbig Ae to Be stars \citep{Donehew2011, Grant2022}. Observations of \citet{Wichittanakom2020} and \citet{Vioque2022} have set a threshold of $\sim$4 M$_{\odot}$ mass for these different accretion regimes, above which magnetospheric accretion might no longer be dominant \citep{Ababakr2017}. Currently, the Boundary Layer mechanism is regarded as an alternative mechanism (described in Sect. \ref{Intro}), however, no observational evidence exists in favor of any proposed mechanism.

Regardless of the accretion mechanism, the Hydrogen transition lines should always be produced by the accretion process, thus they should trace the accretion luminosity provided the outflow contamination is not significant (and provided an hyper-compact \hii  region has not yet formed), therefore new calibrations focused on intermediate- to high-mass protostars have the potential of revealing accurate $L_{\rm line}$ to $L_{\rm acc}$ relations that can provide reliable accretion rates. %

\subsection{Contaminants: Outflows, diffuse emission, and (HC-)\hii  regions} \label{Discussion:outflows}

Since accretion is not the only phenomenon producing \hi lines, we identified all possible contaminants affecting our detections (see Table \ref{tab:hi_lines}). We identified outflows based on the premise that if they are present, they should appear elongated (e.g. Fig. \ref{Fig:G31_outflow_mosaic}). We also tested the presence of diffuse \hi with an annulus around (or very close to) each protostar, and determined whether (hyper-compact) \hii  regions are present from centimeter (cm) data in literature.%

Figure \ref{Fig:G31_outflow_mosaic} demonstrates that \hi lines also trace outflows, in this case toward G31. %
The mosaic displays continuum subtracted maps of G31 for five of the detected \hi lines: \hi 6--5, 8--6, 7--6, 8--7, and 10--7, and includes two known tracers of outflow emission in the bottom row, the H$_2$ 0-0 (S1) and [Fe$\,\textsc{ii}$] emission lines at 17.035 \mum{} and 24.52 \mum{} respectively.  In particular, it becomes clear that \hi 6--5 and \hi 7--6 are tracing extended emission that is consistent with that traced by H$_2$ and [Fe$\,\textsc{ii}$]. This comparison shows that the \hi lines are not seen toward the driving source of G31, but are only coming from its outflow when detected. Ultimately, this implies that, for compact \hi line emission, outflows can still contaminate the observed \hi flux if the beam encompasses spatial scales larger than that of the immediate protostellar envelope. Since this is the case for the MIRI/MRS beam size toward our sources, outflow contamination cannot be fully ruled out for our detections.

Overall, outflows are present in the field of view of all the six regions, either traced in the MIR by \hi lines (G31); H$_2$ vibration lines (all regions); or in the millimeter by SiO (2--1) in G28IRS2 \citep{Beuther2026}, CO in G28P1 (\citealt{Kong2019b}), SiO (2--1) and CO (2--1) in G28S (\citealt{Tan2016, Feng2016,Pillai2019}), SiO (5--4) in IRAS18089 (\citealt{Beuther2004a}), or both SiO (2--1) and HCO$+$(1--0) in IRAS23385 (\citealt{Molinari1998,Gieser2023b}). 
More generally, this also holds for low-mass sources, all of which show outflows and/or jets, yet some fraction comes from accretion (shock conditions studied in \citealt{Navarro2025}). Further work will examine these JOYS outflows through e.g. H$_2$, [Fe$\,\textsc{ii}$] or [S$\,\textsc{i}$] lines, and will quantify ejection rates as a complement to accretion rates. 

Solid evidence for a hyper-compact \hii region exist for IRAS18089mm \citep[e.g.][]{Sridharan2002, Beuther2002c,Zapata2006}, and sources of free free emission are reported toward the millimeter peak of G31 \citep{Cesaroni2010}. In both cases, the cm emission data have angular resolution comparable to our observations. The other regions are also studied at cm wavelengths but no compact sources are detected (see \citealt{Beuther2026} for the G28 subregions; and \citealt{Molinari1998a} for IRAS23385).

\subsection{Methodological caveats} \label{Discussion:Caveats}

Some important sources of uncertainty arise when attempting to quantify accretion rates for embedded, massive protostars. The main ones, ordered by decreasing importance according to our experience here, are four: 

\begin{itemize}

 \item[(1)] The validity of empirical $L_{\rm HI}$ to $L_{\rm acc}$ relations when extended to massive protostars.

 \item[(2)] The disagreement between available extinction curves in literature, especially at wavelengths around the broad silicate absorption feature peaked at 9.7 $\mu$m. 

 \item[(3)] The estimation of $\tau_{9.7}$.
 
 \item[(4)] Stellar parameters: mass and radius of protostellar embryo.
\end{itemize}
 
We elaborate on each point in the following. 
On caveat (1), it remains unclear how valid it is to extrapolate calibrations established for low mass Class II objects, to massive protostars that actually show much higher $L_{\rm line}$. In line with our findings, the works suggesting that the accretion mechanism might vary for higher stellar masses (from $\sim 4 \, \rm M_{\odot}$, Sect. \ref{Discussion:validity-of-tracers}) strongly motivates the development of new calibrations adapted to these higher mass regimes, for which MIRI-MRS would be critical. A feasible approach would be searching for several intermediate- to high-mass sources with known accretion parameters (measured using tracers in the near-IR). Alternatively, contemporaneous measurements of MIR \hi tracers and \Lacc{} can mitigate the uncertainty given by accretion variability. Crucially, the possibility to use \hi lines as tracers of accretion luminosity remains unaltered despite this caveat. 

On caveat (2), the two dust extinction models we implement in this work can provide contrasting factors depending on wavelength, therefore \hi lines are unequally affected and conclusions become harder to extract. The clearest evidence of this is the  factor of $\sim 1.6$ difference that separates the extinction of \hi 7--6 (12.37 $\mu$m) measured using the KP5 and M09 curves (Table \ref{tab:extinction}). More generally, Table \ref{tab:fitted_accretion_lines} shows the two possible luminosities for each line, reflecting how each of them gets affected. From this particular perspective, \hi 6--5 and \hi 8--6 gain more reliability as the dust extinction curves differ less at their wavelengths.

Moving to caveat (3), the fitting results still depend on what absorbing species are considered. Although water ice and silicates seem to be dominant, our fits achieve different levels of accuracy. Similarly, the wavelength ranges provided to the fitter as input also affect the results. Since the main absorption feature at 9.7 \mum{} goes deeper than the noise floor for all our sources, we miss a relevant fraction of both the slope at R2, and the bluer tip of R3. To conclude, incorporating more absorbing species (e.g. HCOOH, CO$_2$) and possibly the PAH emission feature, together with deeper observations of the main silicate feature, has the potential of yielding more accurate fits. Alternatively, Chen et al., in prep., provide a complementary approach to model protostellar spectra.

Caveat (4) arises on the last step, affecting accretion rates only.  Indeed stellar parameters are frequently unknown for the most embedded stages as their envelopes tend to be more massive. However, promising progress has been made to estimate embedded protostellar masses using NOEMA or ALMA from molecular line kinematics of young disks. Although mostly done for sources closer than $\sim 400 $ pc \citep{Lee2017, Tobin2024} where masses reach up to 2.6~M$_{\odot}$ \citep{Ohashi2023}, measurements on massive protostars have been achieved at distances of 2~kpc, pointing to masses up to 25~M$_{\odot}$ \citep{Ahmadi2023}.  New observations with existing facilities can potentially clear the mist around the stellar parameters toward the (usually kpc-distant) high mass protostars.

In addition to these caveats, we emphasize the uncertainty that a difficult-to-quantify fraction of the compact emission may come from outflows in more cases than those we report (see discussion in Sect. \ref{Discussion:outflows}).

\section{Conclusions} \label{Conclusions}

We performed the first accretion-focused JWST study for a sample of high-mass star-forming regions, extending from the better-known low-mass Class II objects to explore the less-understood massive protostellar objects. We searched for compact emission from Hydrogen transitions, we unveiled what the stumbling blocks can be for any similar study, and therefore showed what the prospects can be, given current instruments, of actually measuring accretion rates onto massive protostars. Our conclusions are the following:

\begin{itemize}

\item[1. ]
Hydrogen emission lines in the MIR can now be detected systematically using MIRI/JWST, although much attention should be invested in understanding what processes are actually driving the emission. Besides accretion, outflows, hyper-compact \hii regions, and environmental diffuse gas can produce \hi emission lines. 

\item[2. ]

Three emission lines of Hydrogen stand out because of their detection rate, which places them as the best candidate tracers of protostellar accretion for massive protostars. These lines are \hi 7--6, \hi 6--5, and \hi 8--6.%

\item[3. ]
Our accretion luminosity estimations carry large uncertainties, which limits the informative value of the accretion rates. For that reason and because the accretion process is believed to change for stars more massive than $\sim 4 $ M$_{\odot}$, this work is revealing the need to develop \Lline{}- to \Lacc{}-calibrations focused on higher mass objects. For such a calibration to yield physical \Lacc{} for intermediate- to high-mass protostars given typical \Lbol{} values, we anticipate a shallower growth of $L_{\rm acc}$ with $L_{\rm line}$ compared to what has been derived for low-mass T-Tauri stars.

\item[4. ]
We emphasize (and largely describe) throughout this work the role of  uncertainties propagated in each step. Importantly, our approach to estimate extinctions $A_{\rm K}$ (see \citealt{Gieser2026subm}) represents a step forward, because it offers solid constraints for at least some of our sources. Ultimately, this brings us closer to achieve systematic characterizations of extincted emission due to massive envelopes.

\end{itemize}

Accretion rates in high-mass protostars remain challenging, however this work is contributing to outline the limitations of current work and possible pathways, given the potential of JWST, to better understand accretion processes in high-mass protostars.

\begin{acknowledgements}
This work is based on observations made with the NASA/ESA/CSA James Webb Space Telescope. The data were obtained from the Mikulski Archive for Space Telescopes at the Space Telescope Science Institute, which is operated by the Association of Universities for Research in Astronomy, Inc., under NASA contract NAS 5-03127 for JWST. These observations are associated with program 1290.
This work is based in part on observations made with the Spitzer Space Telescope, which was operated by the Jet Propulsion Laboratory, California Institute of Technology under a contract with NASA.
This paper makes use of the following ALMA data: ADS/JAO.ALMA\# 2017.1.00101.S, 2013.1.00489.S, 2019.1.00195.L and 2011.0.00429.S. ALMA is a partnership of ESO (representing its member states), NSF (USA) and NINS (Japan), together with NRC (Canada), NSTC and ASIAA (Taiwan), and KASI (Republic of Korea), in cooperation with the Republic of Chile. The Joint ALMA Observatory is operated by ESO, AUI/NRAO and NAOJ.
The Submillimeter Array is a joint project between the Smithsonian Astrophysical Observatory and the Academia Sinica Institute of Astronomy and Astrophysics and is funded by the Smithsonian Institution and the Academia Sinica.
A.C.G. acknowledges support from PRIN-MUR 2022 20228JPA3A “The path to star and planet formation in the JWST era (PATH)” funded by NextGeneration EU and by INAF-GoG 2022 “NIR-dark Accretion Outbursts in Massive Young stellar objects (NAOMY)” and Large Gran INAF-2024 “Spectral Key fea-tures of Young stellar objects: Wind-Accretion LinKs Explored in the infraRed (SKYWALKER)”. V.J.M.L.G. acknowledges support by the Unidad de Excelencia María de Maeztu program number CEX2020-001058-M. V.J.M.L.G. acknowledges support by the European Research Council (ERC) under the European Union’s Horizon 2020 research and innovation program (grant agreement No. 101098309 - PEBBLES).

\end{acknowledgements}


\bibliographystyle{aa} 
\bibliography{bib/ref} 

-------------------------------------------------------------------
\appendix

\section{Detailed \hi line detections} \label{Appendix:detailed-HI}
In the following we dive deeper into each line detection, highlighting important aspects of their derived parameters, including caveats, with the purpose of enabling further interpretations of how much they can be used to constrain accretion parameters toward massive protostars (Sect. \ref{Discussion:validity-of-tracers}).

\textbf{Humphreys $\alpha$ (\hi 7$-$6):} it is the most frequently detected line in the sample of \cite{Rigliaco15}, which remains true in our smaller sample despite Hu $\alpha$ being strongly affected by the silicate absorption feature at 9.7 $\mu$m.
Among our four detections, that of G28IRS2 stands out with SNR = 7.4 (Fig. \ref{Fig:map_spec1}, third row), and with a protostar emission about 1.6 times brighter than that of the local background, making it a relatively secure detection. In this case, the neighbor line HI(11$-8$) is never detected.
G28P1 A and B show tentative detections, but G28P1 B is detected only 1\% above the background ($L_{\rm proto}/L_{\rm bkg}$=1.01), hence it is the least secure detection of all. 
Lastly, IRAS23385 A is fitted with SNR of 2.5, while its local background shows no emission. 
We can compare the latter result toward IRAS23385 A  with that of \citet{Beuther23}, who report a tentative detection with SNR $\sim 3$. There are key differences because they used an aperture of 2$^{\prime \prime}$ radius that encompass both IRAS23385 A and B, but since IRAS23385 A is the sole driver of compact ionization we centered the aperture at this protostar only, adopting a smaller aperture of $0.^{\prime \prime}5$ (same for all our regions). Their larger aperture adds emission from \hi 11--8, which is particularly seen diffuse in IRAS23385, and whose flux has to be summed to that of \hi~7--6 when applying the calibration of \citet{Rigliaco15} (as pointed out in \citealt{Franceschi24}). All of these explain our lower SNR of $2.5$. Our integrated line luminosity is lower, however, as we measured a 2.2 times higher A$_K$ extinction (likely due to our smaller aperture size), our $L_{\rm acc}$ becomes $\sim 5$ times higher (using same dust extinction model). Overall, our lower SNR reduces the certainty of the \hi 7--6 detection, however, the copious \hi $6-5$ and \hi $8-6$ compact emission confirms the presence of ionizing activity at the IRAS23385 A source.

\textbf{Pfund $\alpha$ (\hi 6$-5$):} It is detected in two protostars with SNR~$>4.5$ and significant $L_{\rm proto}/L_{\rm bkg}$ ratio (Table \ref{tab:fitted_accretion_lines}). The line is also detected as an outflow in G31 due to its very extended emission, and toward IRAS18089mm, which is probably an HC-H$\,\textsc{ii}$ region since 0.7, 1.3, and 3.6 cm emission is detected with spectral index of 0.58 \citep{Sridharan2002, Beuther2004a, Zapata2006}, indicative of a slightly compact, optically thick region with free free emission.

\textbf{Humphreys $\beta$ (\hi 8$-6$):} It is detected in three protostars as compact emission, with no diffuse environment contamination in two cases.%
Two line detections deserve small comments regarding their validity. The G28IRS2 detection has a high SNR of 8.7, however its Gaussian peak appears slightly redshifted by about 1.5 channels (i.e. $\sim 48$ km$\, \rm s^{-1}$), which does not seem to be a detector issue because the line at the background appears where expected given the G28 $v_{\rm lsr}$ as shown in Fig. \ref{Fig:map_spec1} (second row, lower inset). The second gaussian component appears two channels blueshifted (i.e. -64 km$\, \rm s^{-1}$) with respect to the \hi 11--7 line (upper inset), and it is less clear that it actually corresponds to this \hi transition. 
Despite this note, we assume this is a true detection and we include it in Table \ref{tab:fitted_accretion_lines}.
On the other hand, the integrated line map of IRAS23385 A in Fig. \ref{Fig:map_spec2} shows some extended structure that may resemble that of outflows, but none of that emission reaches a SNR of 2.5, therefore we treat it as a compact detection produced by accretion.
%

\textbf{\hi 10$-7$:} We have two tentative detections with no diffuse contamination nor extended emission associated to outflows. %
However, we judged them as dubious because that of G28P1B appears two channels blueshifted ($-64$ km$\, \rm s^{-1}$) given its $v_{\rm lsr \,}$, and that of IRAS18089C has a narrow baseline that affects the noise measurement (Appendix \ref{Appendix:contsubIRAS18089}). For these reasons they are flagged in Tables \ref{tab:hi_lines} and \ref{tab:fitted_accretion_lines}, and excluded from the analyses of Sect. \ref{Discussion}.

\subsection{Studied \hi transitions}

\begin{table}[h!]
\centering
\caption{All \hi transitions considered in this work.}
\begin{tabular}{ll}
\hline
\textbf{\hi Transition} & \textbf{Wavelength [$\mu$m]} \\
\hline
10--6 & 5.1286575 \\
9--6  & 5.9082134 \\
12--7 & 6.7719906 \\
6--5  & 7.4598577 \\
8--6  & 7.5024932 \\
11--7 & 7.5081049 \\
10--7 & 8.7600642 \\
13--8 & 9.3920177 \\
12--8 & 10.503499 \\
9--7  & 11.308695 \\
7--6  & 12.371898 \\
11--8 & 12.387168 \\
14--9 & 12.587077 \\
13--9 & 14.183084 \\
10--8 & 16.209104 \\
15--10& 16.411724 \\
12--9 & 16.880628 \\
14--10& 18.615151 \\
8--7  & 19.061898 \\
16--11& 20.920562 \\
13--10& 22.331574 \\
11--9 & 22.340457 \\
15--11& 23.868016 \\
17--12& 26.168218 \\
9--8  & 27.803379 \\
\hline
\end{tabular}
\label{tab:HI_wavelengths}
\end{table}

\begin{figure*}[h!]
\centering
\includegraphics[width=0.99\linewidth]{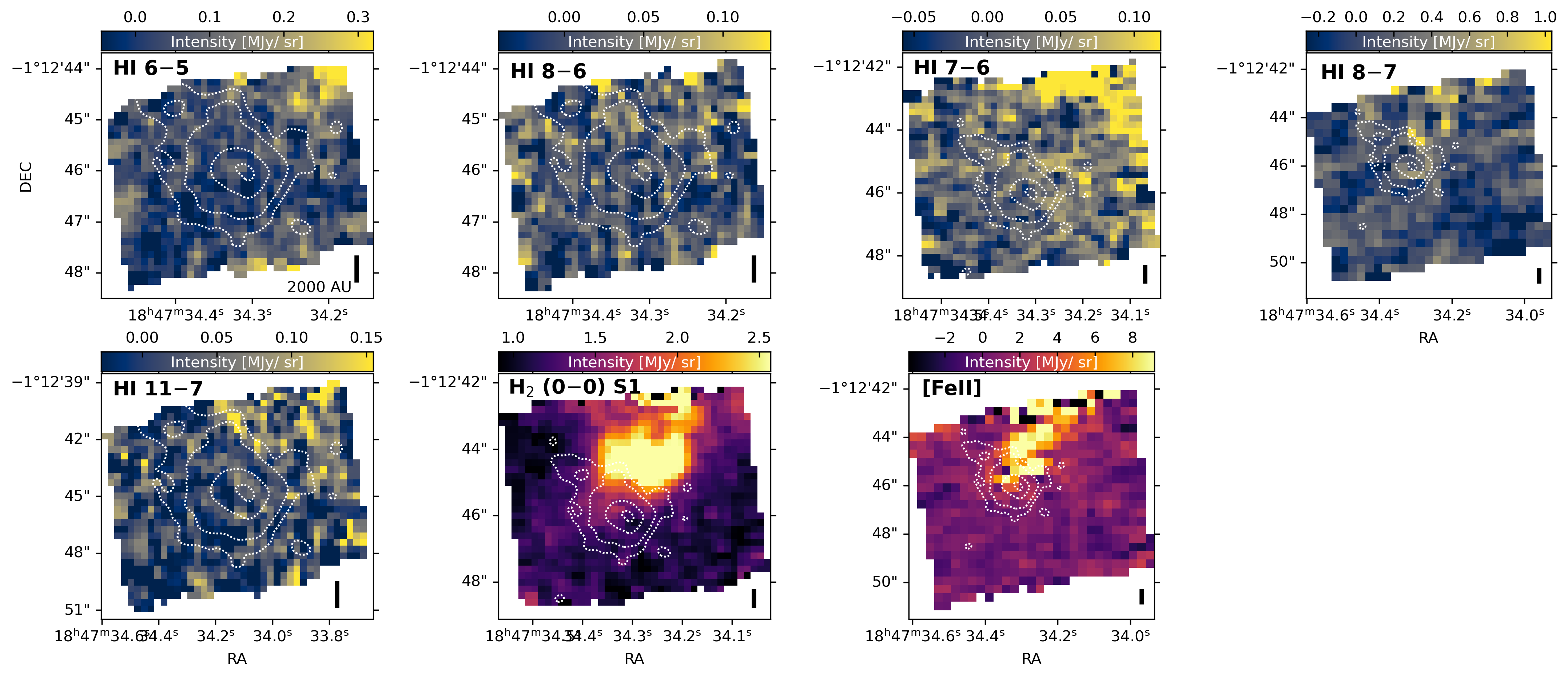}
\caption{Integrated map of \hi lines and the outflow tracer lines H$_2$ (17.03 \mum{}) and [Fe$\,\textsc{ii}$] (24.52 \mum{})  in the G31 region. White contours show ALMA 1.3 mm continuum. The contrasting emission of the outflow tracers demonstrate that, for this region, the \hi lines are mainly produced along the outflow (see Table \ref{tab:hi_lines}). }
\label{Fig:G31_outflow_mosaic}
\end{figure*}

\begin{figure*}

\includegraphics[width=0.79\linewidth]{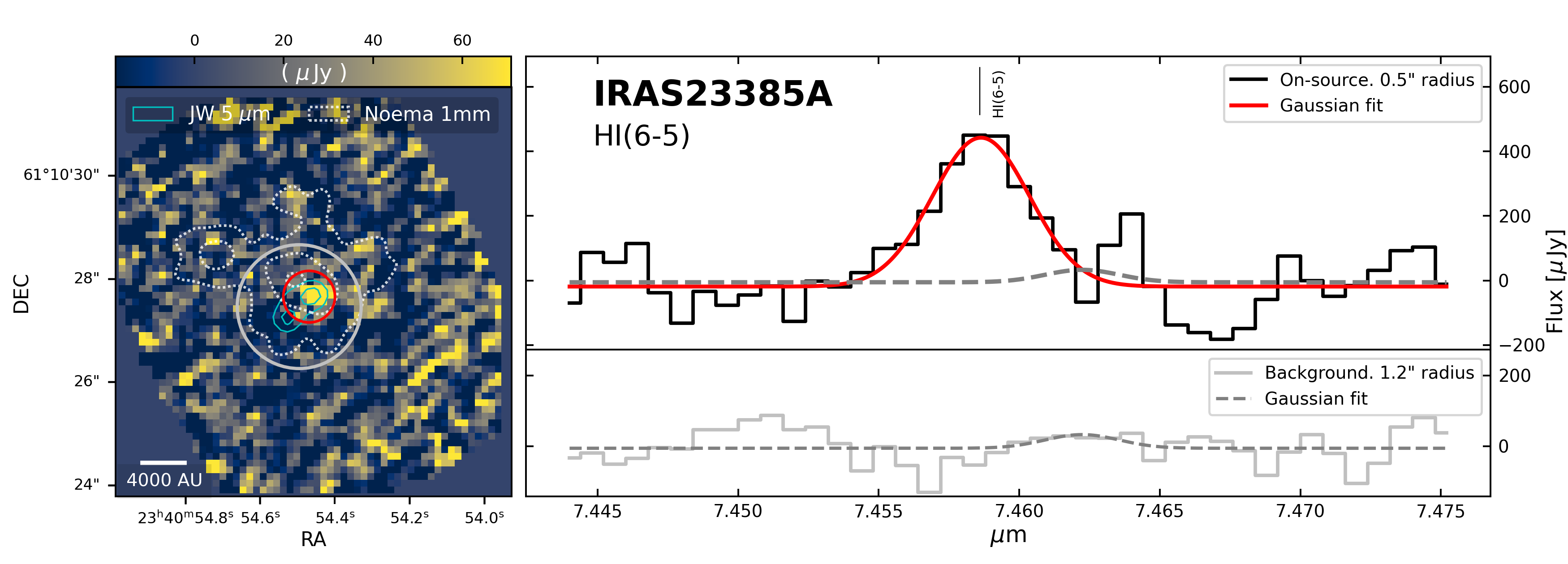}
\includegraphics[width=0.79\linewidth]{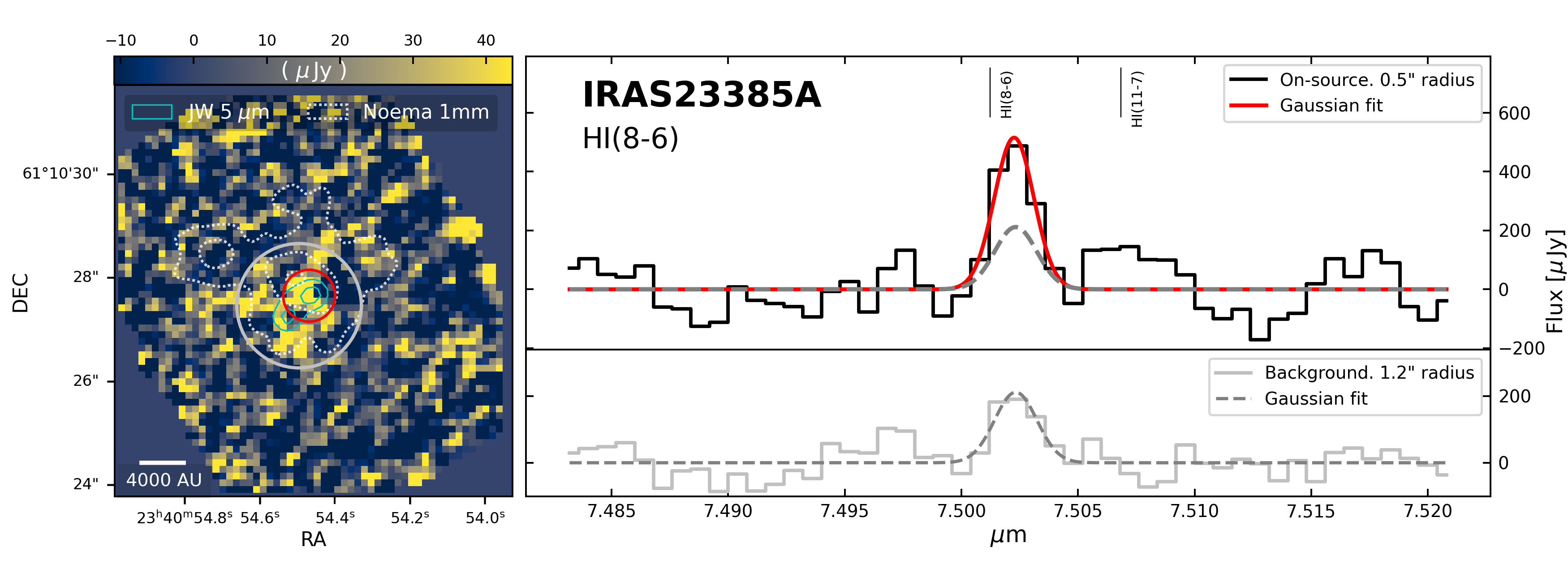}
\includegraphics[width=0.79\linewidth]{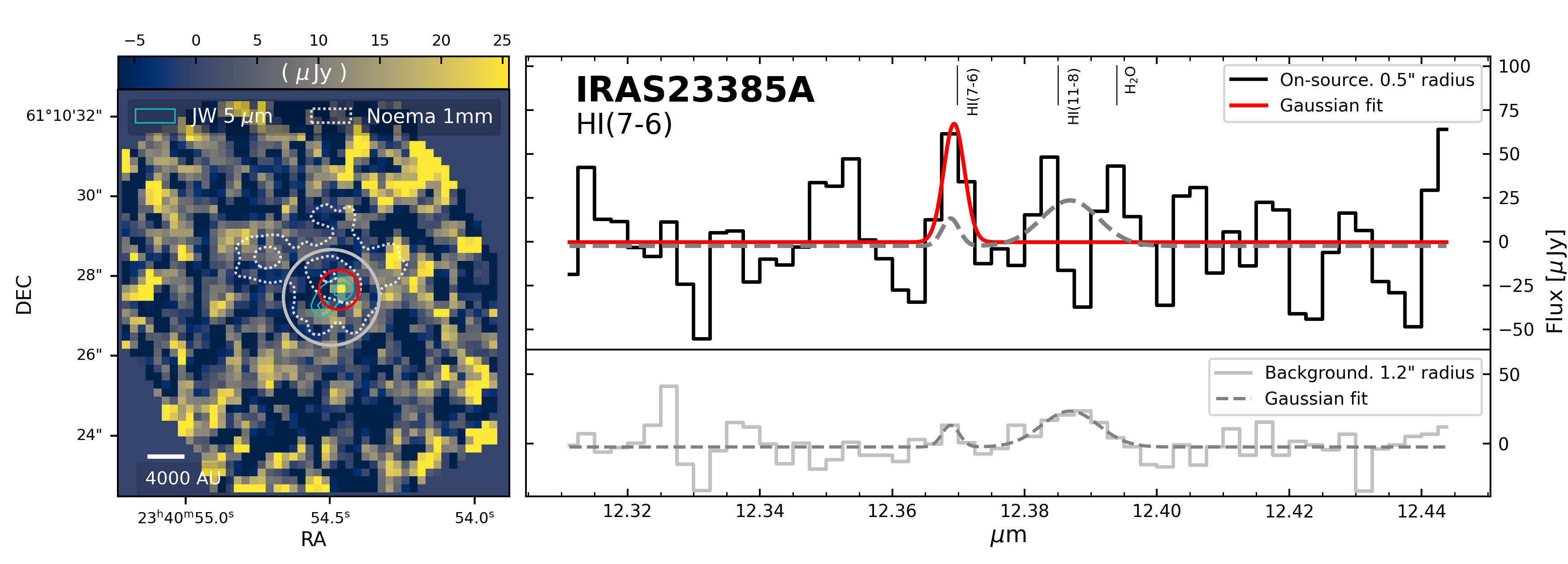}
\includegraphics[width=0.79\linewidth]{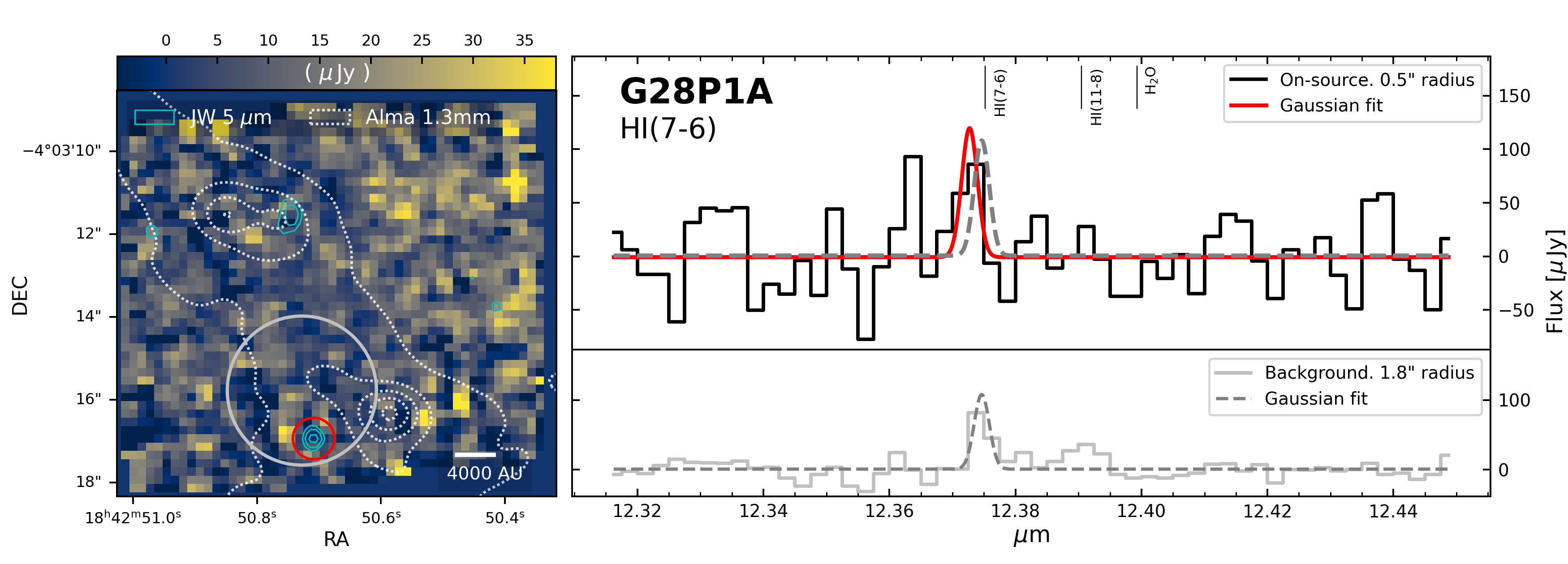}

\caption{Continuation of Fig. \ref{Fig:map_spec1} for IRAS23385 (\hi transitions 6--5, 8--6, 7--6) and G28P1A (\hi 7--6).\\
Left: line moment 0 maps. The channels used to integrate the emission are those encompassed by the line Gaussian fits in the right panels. Red apertures enclose the map area used to extract the spectra at the protostars shown in right panels (upper insets), while the broader grey aperture (minus the central red aperture) is used to extract the spectra of the local background (right panels, lower insets). Right: Gaussian curves fit the line emission for both protostar (red, upper inset) and backgrounds (grey, lower insets). We reproduce the Gaussian fits of the background in upper insets to allow comparisons with the protostar Gaussian fit.}
\label{Fig:map_spec2}
\end{figure*}

\begin{figure*}
\includegraphics[width=0.79\linewidth]{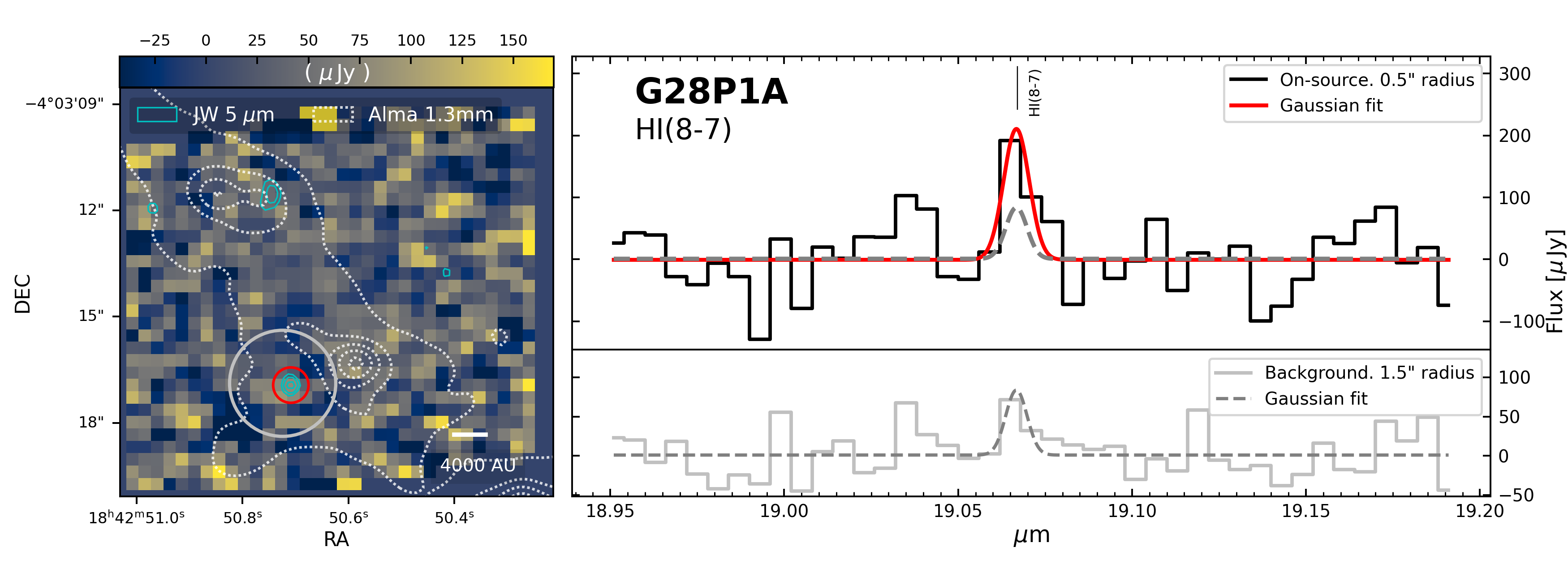}
\includegraphics[width=0.79\linewidth]{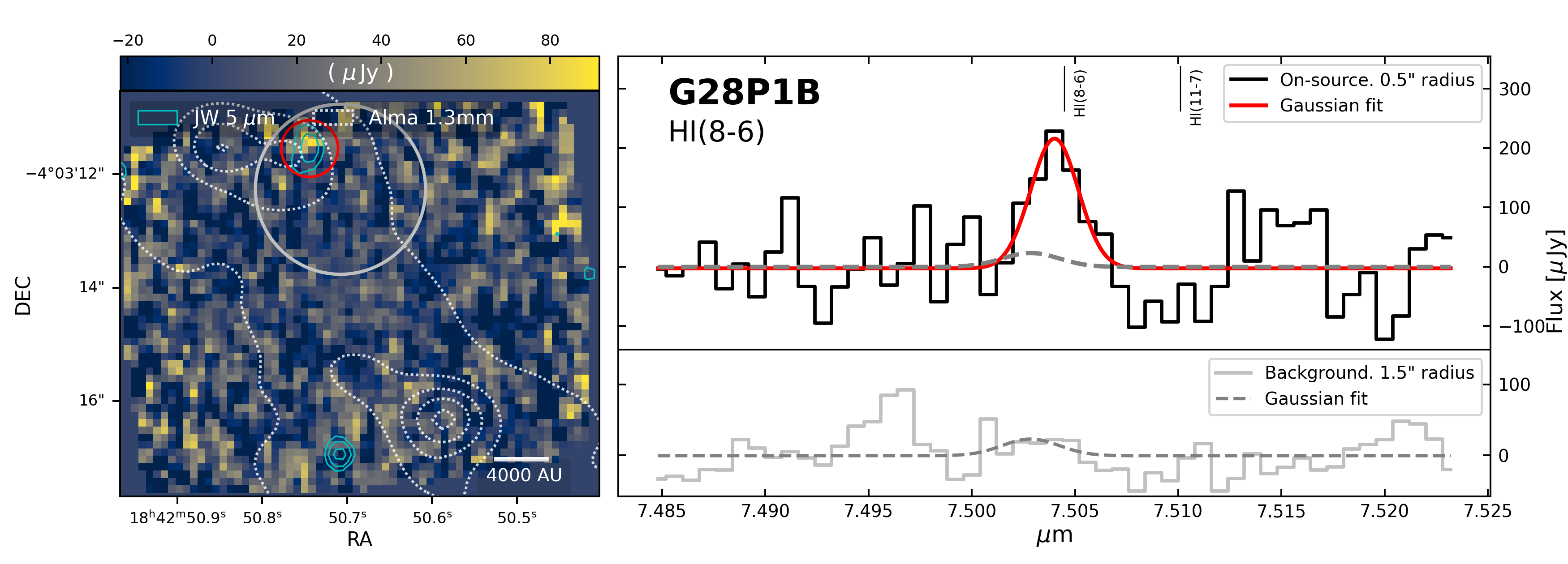}
\includegraphics[width=0.79\linewidth]{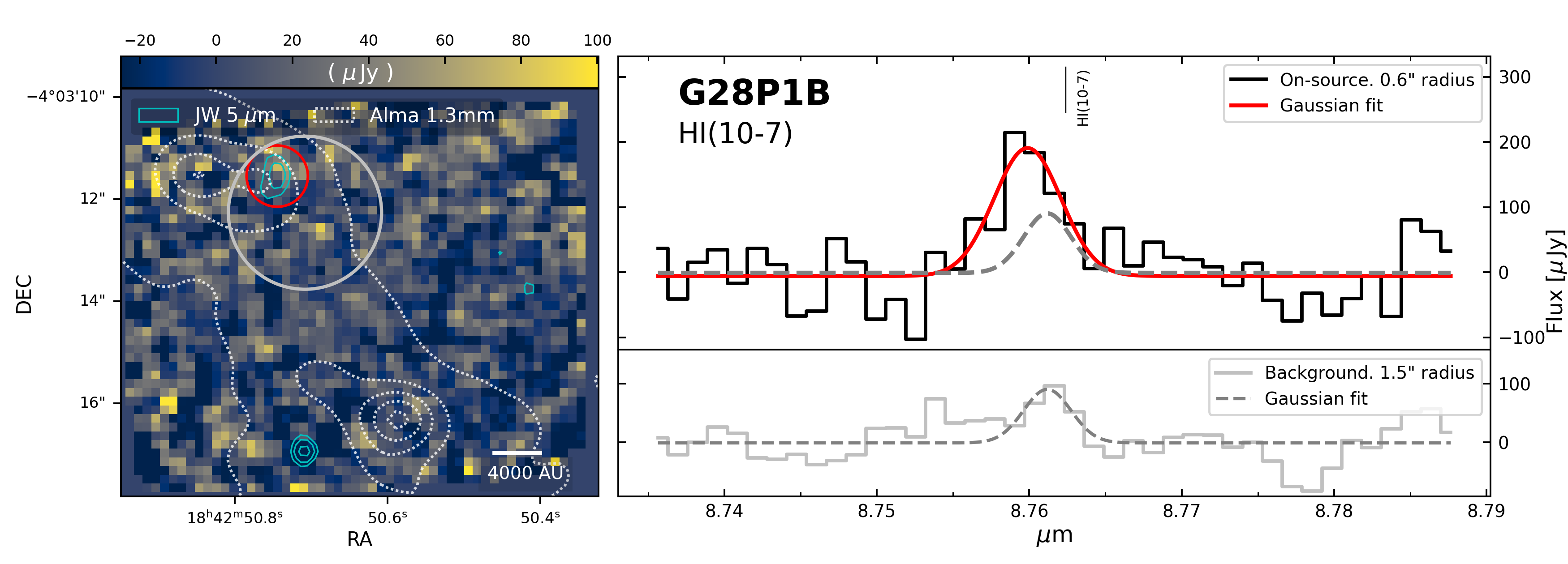}
\includegraphics[width=0.79\linewidth]{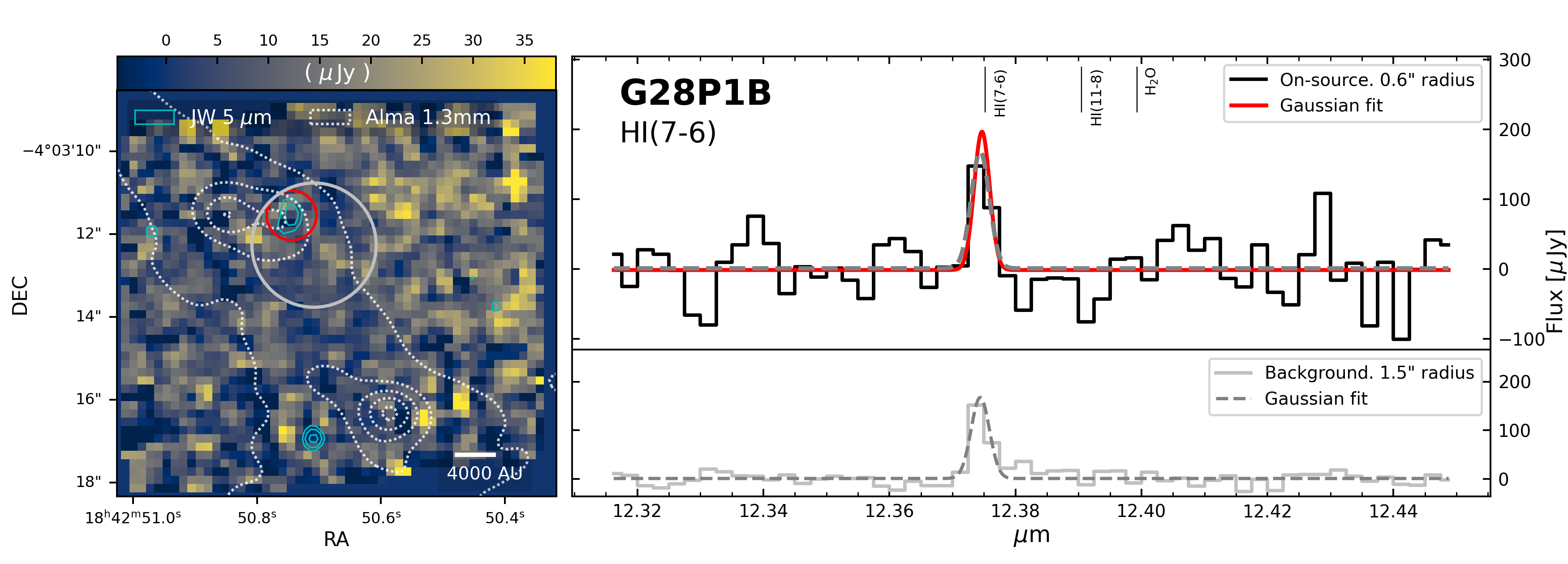}

\caption{Continuation of Fig. \ref{Fig:map_spec1} for G28P1 A and B. }
\label{Fig:map_spec3}
\end{figure*}

\begin{figure*}
\includegraphics[width=0.79\linewidth]{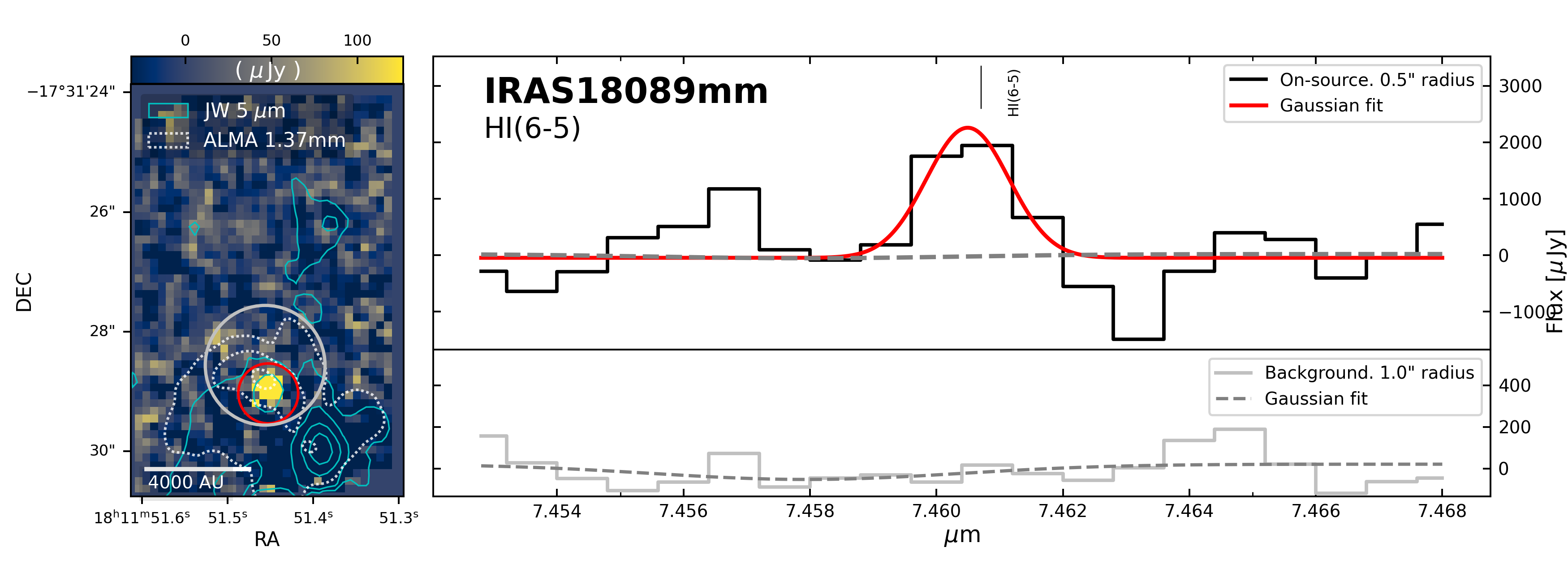}
\includegraphics[width=0.79\linewidth]{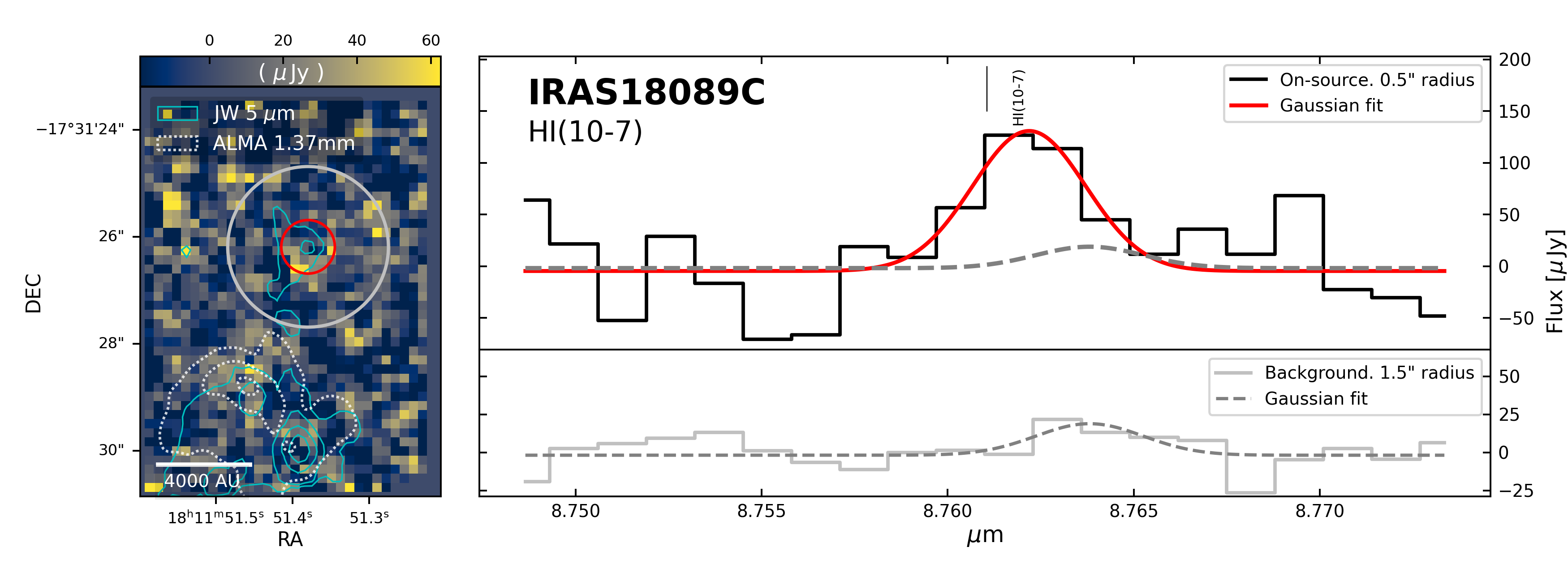}
\caption{Continuation of Fig. \ref{Fig:map_spec1} for IRAS18089 A and mm. }
\label{Fig:map_spec4}

\end{figure*}

\begin{figure*}[h!]
\includegraphics[width=0.8\linewidth]{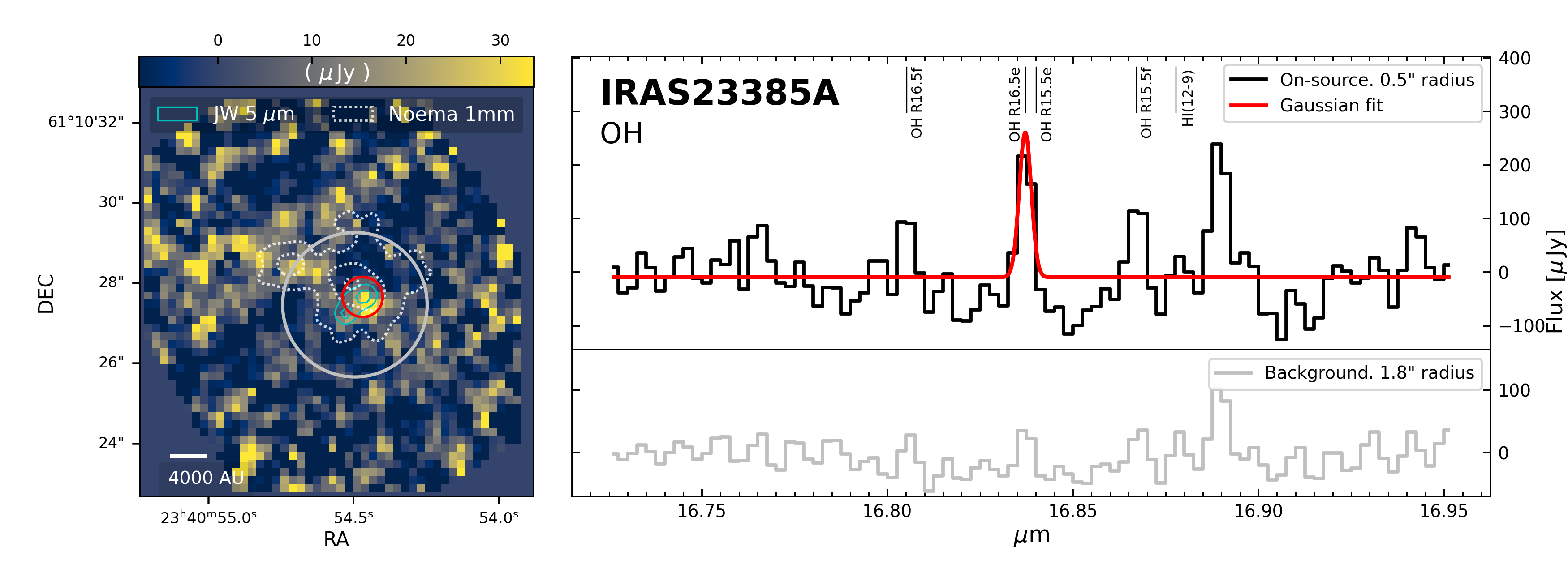}
\includegraphics[width=0.8\linewidth]{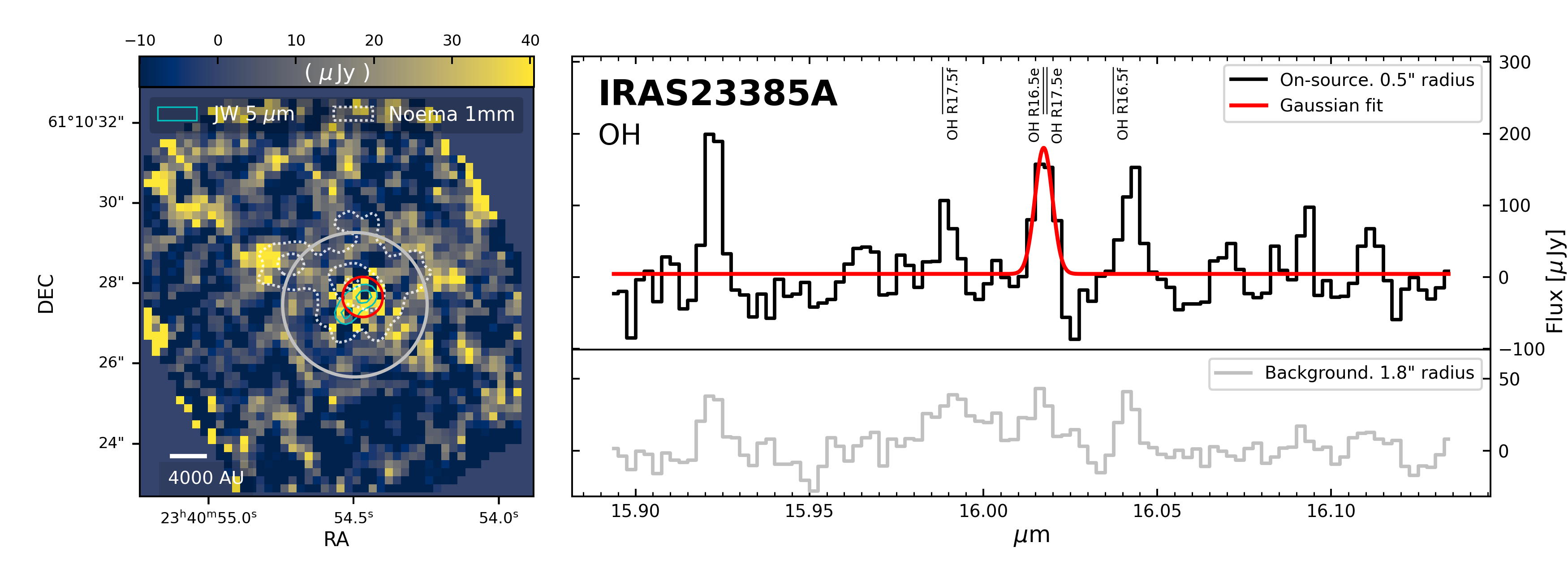}
\caption{Detection of OH emission in IRAS23385 A. We include a single-component Gaussian fit to the central line, which is a narrow blend of the two central lines in the OH quartet. }
\label{Fig:OH_IRAS23385A}

\end{figure*}

\begin{figure*}[h!]
\includegraphics[width=0.76\linewidth]{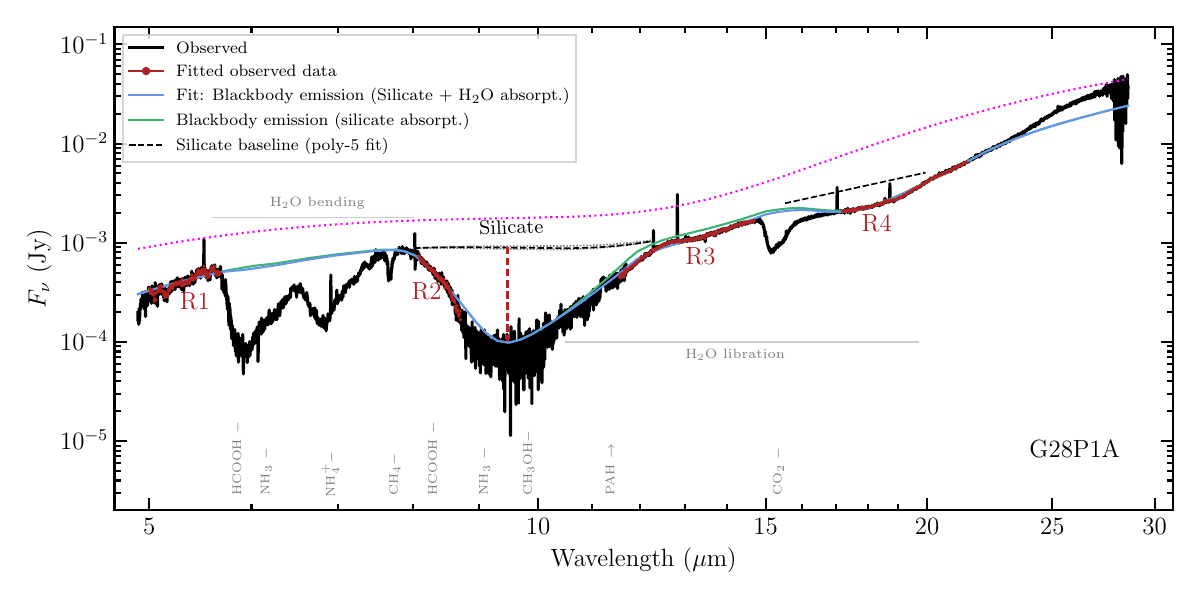}
\includegraphics[width=0.76\linewidth]{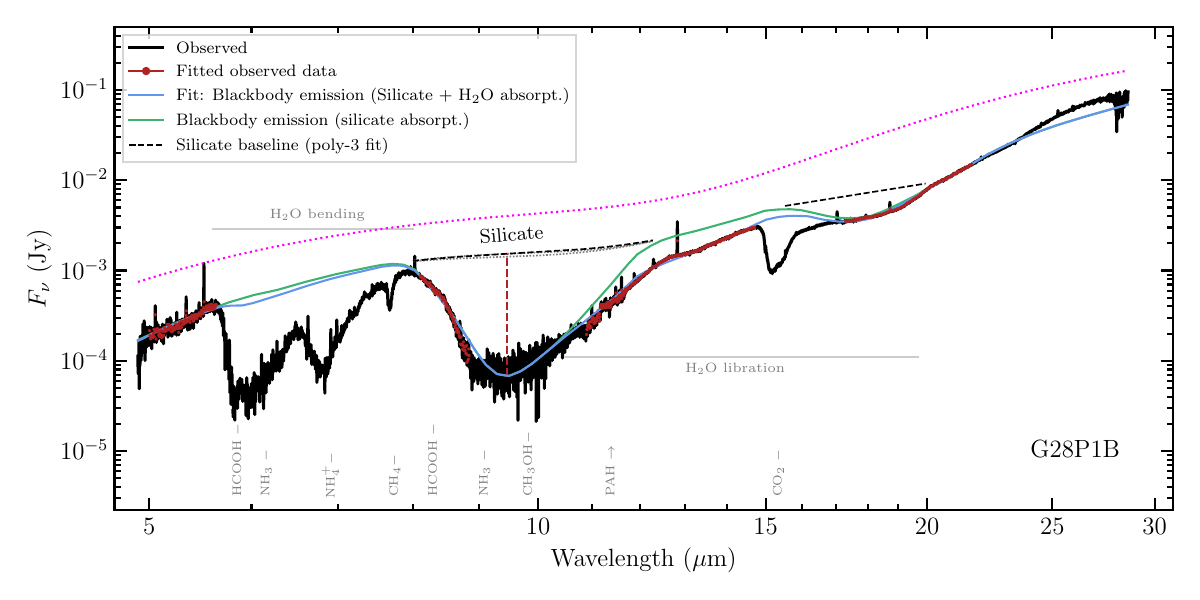}
\includegraphics[width=0.76\linewidth]{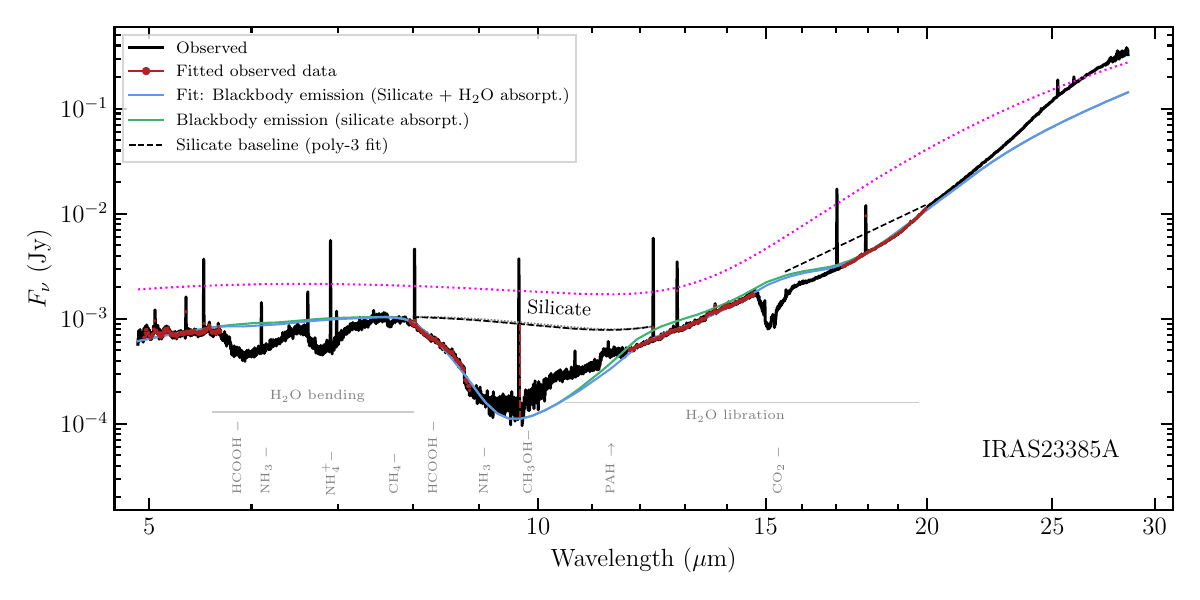}
\caption{Continuation of Figure \ref{Fig:silicate_fit_G28IRS2}. Full MIRI spectrum of the labeled protostars (black). It shows several absorption features, where the broader one at 9.7 $\mu$m is produced by silicate grains. Two blackbody components absorbed by silicates and H$_2$O are fitted to reproduce the overall continuum. The blue curve represents the fit obtained by considering those spectral ranges that are predominantly absorbed by silicates and H$_2$O (in red, also indicated by the R1-R4 anchor ranges in top panel). The green curve shows the modeled blackbody emission only absorbed by silicates, while the dotted magenta curve shows the modeled blackbody emission unaffected by absorption. The black dashed curve above the main silicate absorption feature at 9.7 \mum{} is an interpolation of the silicate-absorbed continuum from a third-order polynomial fit. Here, the red vertical line highlights the resulting depth of the feature that determines $\tau_{9.7}$, whose uncertainty is given by the other two polynomial fits passing above and below (grey curves). The black dashed line at R4 indicates the location of the secondary silicate absorption feature.}
\label{Fig:sillicate_fits2}
\end{figure*}

\begin{figure*}[h!]
\includegraphics[width=0.76\linewidth]{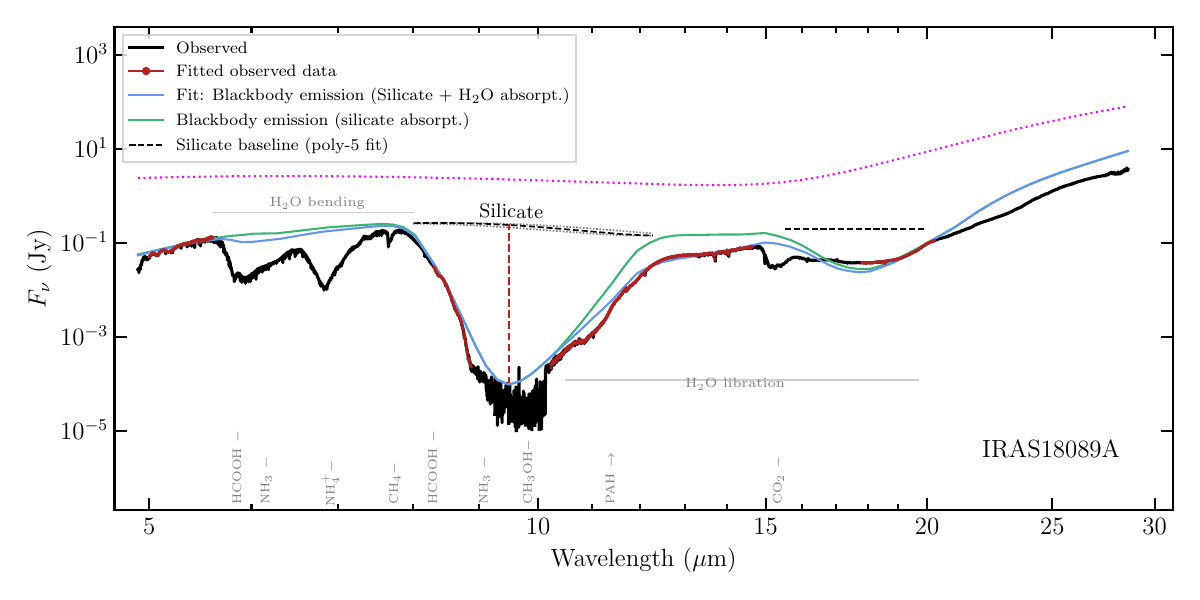}
\includegraphics[width=0.76\linewidth]{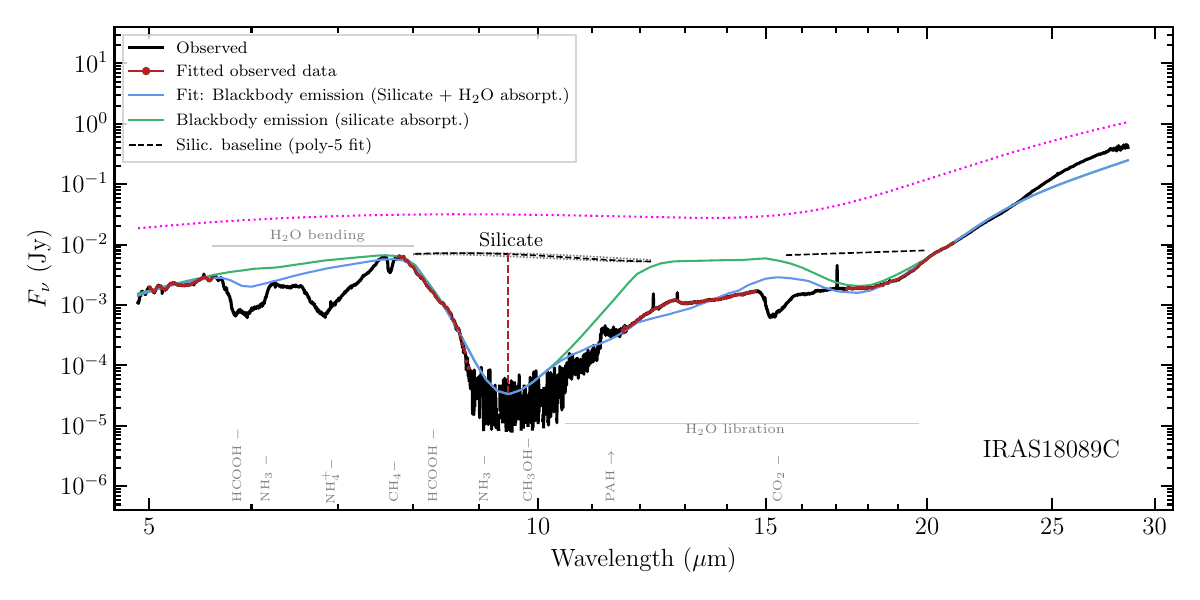}
\caption{Continuation of Figure \ref{Fig:silicate_fit_G28IRS2}. For IRAS18089 A and C we masked the very deep absorption features at 5.9 and 6.9 $\mu$m as it shows mostly noise. }
\label{Fig:sillicate_fits3}
\end{figure*}

\begin{table*}[h!]
\caption{Fitted parameters from the spectrum of each IR source.}
\begin{tabular}{lrlrlccllll}
\midrule
Source & $T_1$  & $\Omega_1$ & $T_2$ & $\Omega_2$ & $T_3$ & $\Omega_3$ &  f$_{\rm \,Olivine}$  & f$_{\rm \, Pyroxene}$ & f$_{\rm \,H_2O}$  (15K) & f$_{\rm \,H_2O}$ (75K) \\
  &[K]&[sr]&[K]&[sr]&[K]&[sr]& [g$\, \rm cm^{-2}$] &[g$\, \rm cm^{-2}$]&&
\\
\midrule
G28IRS2      & 7731 & 1.7E-21 & 189 & 1.5E-16 & 44 & 1.2E-10 & 1.6E-39 & 9.2E-04 & 9.0E+00 & 1.4E-139 \\
IRAS23385A   & 776  & 2.4E-19 & 72  & 1.7E-13 & -- & -- & 1.8E-04 & 5.1E-04 & 7.4E-01 & 1.6E-15 \\
G28P1A       & 543  & 5.7E-19 & 97  & 4.7E-15 & -- & -- & 1.5E-10 & 6.4E-04 & 7.5E-01 & 3.8E-05 \\
G28P1B       & 391  & 4.1E-18 & 91  & 2.4E-14 & -- & -- & 1.4E-19 & 9.0E-04 & 2.5E+00 & 5.7E-24 \\
IRAS18089A   & 800  & 2.7E-16 & 64  & 1.3E-10 & -- & -- & 1.6E-04 & 2.1E-03 & 4.9E+00 & 2.5E-41 \\
IRAS18089C   & 582  & 8.5E-18 & 64  & 1.7E-12 & -- & -- & 2.0E-39 & 1.5E-03 & 8.2E+00 & 1.5E-23 \\
\midrule
\end{tabular}
\label{tab:extinction_fitting_parameters}
\end{table*}

\section{Post-Processing Steps} \label{Appendix-A}

\subsection{Astrometric correction} \label{Appendix:Astrometric-correction}

To correct the astrometry of all our data cubes we retrieved from the MAST portal (already reduced) parallel MIRI images and point-source catalogues for five of our sources $-$ only for G28P1 these level 3 data products were not available. We found Gaia sources matching catalogued MIRI sources for only three of the regions. Within 0.5" they are G31 (10 matches), G29IRS2 (4 matches), and G28S (4 matches). Their relative astrometric shifts are $\Delta$RA = 100.7 and $\Delta$DEC = 27.0 mas for G31; $\Delta$RA = 158.0 and $\Delta$DEC = 15.26 mas for G28IRS2; and $\Delta$RA = 141.9 and $\Delta$DEC = 61.4 mas for G28S.  Despite the small number of Gaia-MIRI counterparts, their shifts in RA and DEC are consistent among the three regions, confirming that the astrometric difference in the pointing of JWST to Gaia is in fact systematic.

We applied the individual shifts we found to equally correct each corresponding region. Particular cases are the following.  G28P1 (the only region without image in MAST) is located in the same molecular cloud as G28IRS2 and G28S and it was observed on the same run within a few hours difference. Hence, we assumed they are equally affected by the JWST pointing and we assigned to it a shift that is the average between the former two, namely 150 and 38 mas in RA  and DEC respectively. For IRAS23385 we used the correction that was already obtained in \citet{Beuther23}, where they used again the MIRI images at 15 $\mu$m to compare $-$ uncatalogued $-$ MIRI point sources to Gaia (corresponding 15 $\mu$m image in \citealt{van_Dishoeck2025}, Fig. C.1). Finally, for IRAS18089 we tried the later approach, but we only found one Gaia source that matched a MIRI point source. Together with the fact that the corresponding shift was very small (less than 100 mas) we did not apply a pointing correction to IRAS18089.

\subsection{Continuum subtraction: the IRAS18089 case} \label{Appendix:contsubIRAS18089}

We describe first the issues we encountered when creating continuum-subtracted maps for IRAS18089, and secondly the extra steps we applied to mitigate them. 
The very high contrast in brightness between IR-source and background pixels in IRAS18089 produces, at the IR bright pixels, significant flux excess (or flux loss) after the continuum subtraction due to the $-$ unavoidable although small $-$ difference of the polynomial fit relative to the real continuum. In other words, at the brightest pixels the flux difference of the real continuum and the misfitted polynomial will still be much higher than the background noise level, making those pixels appear very bright (or very dark) on our maps depending on whether the fit missed the continuum from below (or above). This in turn produces "chequered" patterns in the integrated intensity maps that make them difficult to analyze.%

\begin{figure*}[h!]
\includegraphics[width=.9\columnwidth]{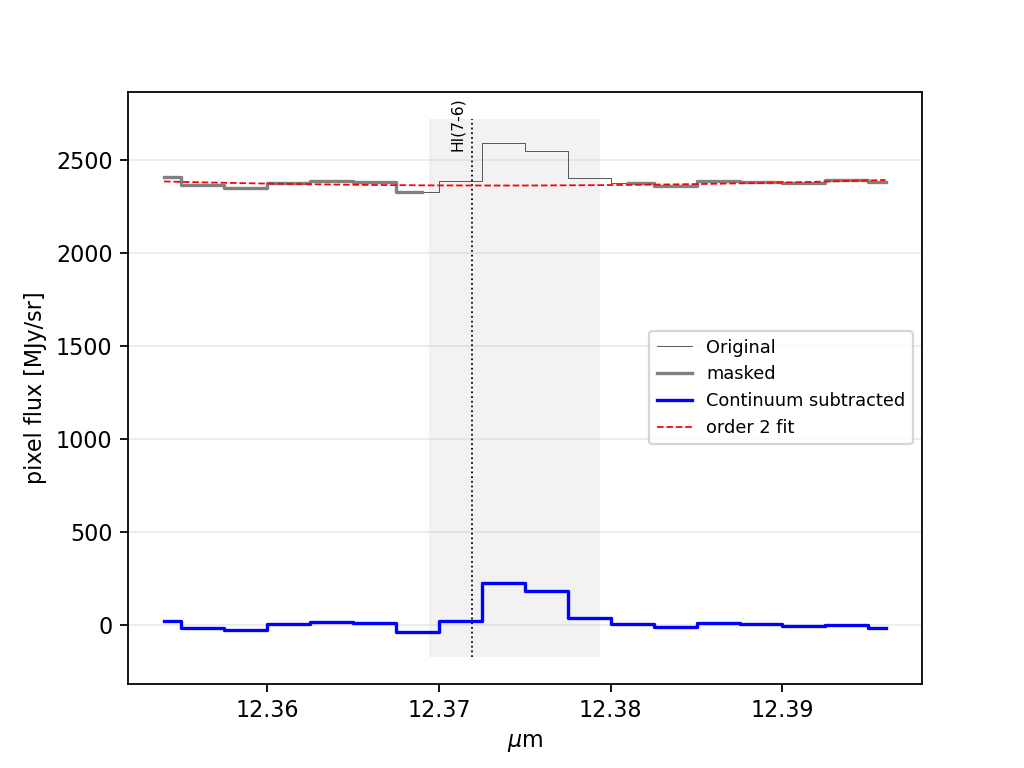}
\includegraphics[width=.9\columnwidth]{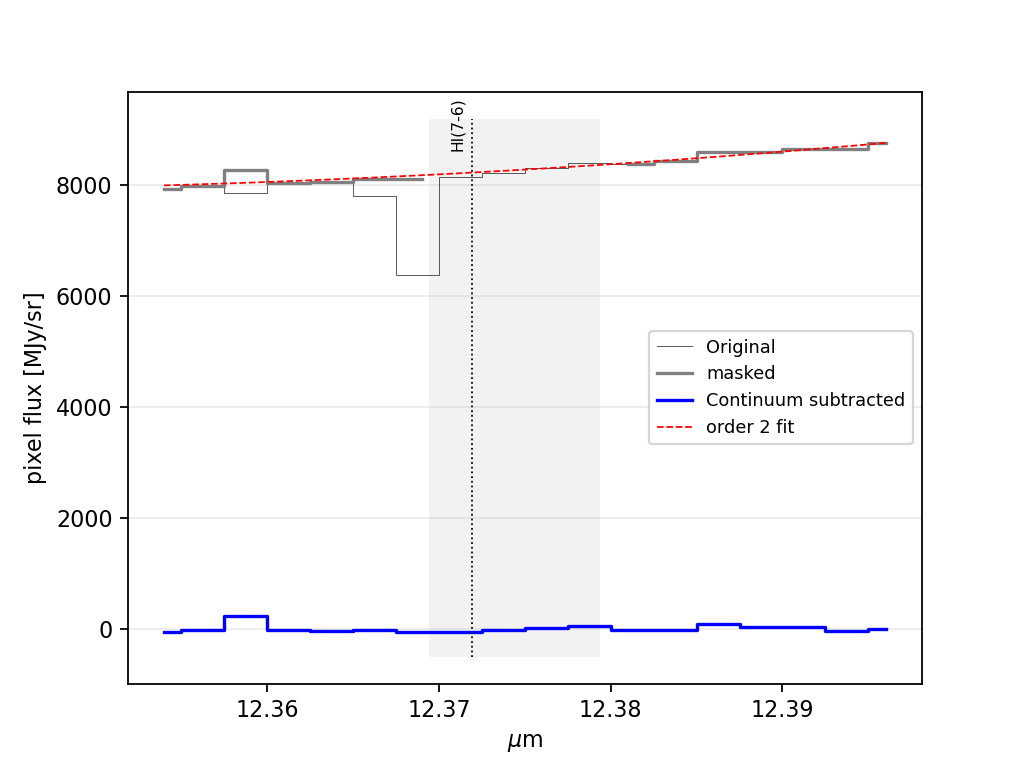}
\caption{Examples of continuum subtraction on two adjacent, high-flux pixels at the IR source A of IRAS18089 that show contrasting behaviors. The  pixel spectrum of the left panel displays the Humphreys alpha line, while the spectrum on the right panel shows a deep artifact that we masked prior to the fitting (red line). In this case the artifact consists of two contiguous, negative channels, where the deeper one is capable of deviating the polynomial fit significantly. The shaded area highlights the spectral range that was excluded when fitting the continuum. The left panel is, at the same time, a good representation of how the continuum subtraction works for the remaining five regions as they do not show negative artifacts. %
The vertical lines show the nominal wavelength of the Hu$\, \alpha$ line.  }
\label{Fig:contsub-IRAS18089}

\end{figure*}

To maximize the fit precision we narrowed the spectral range used to fit the polynomials, leaving only seven channels on each side of the masked spectrum, although at the expense of a more accurate noise estimate. Figure \ref{Fig:contsub-IRAS18089} illustrates how the continuum subtraction works. Pixel spectra tend to follow flat, curved, or wiggled shapes, hence fitting simultaneously three polynomials (order 1, 2, and 3) to each pixel ensures that each case will be modeled.
On a parallel track, we encountered single channels with relevant flux looses. They appear in several bright pixels at the IRAS18089 source locations, making the continuum subtraction imprecise (see right panel in Fig. \ref{Fig:contsub-IRAS18089}). As these single-channel bowls are artifacts, we masked them from our data cubes on a 5 sigma detection basis prior to fitting the continuum, while assigning them instead the average value of the few neighbor pixel channels.  %
Lastly, we applied a 1D fringe correction using the python package jwst.residual\_fringe.utils, although its correcting effect is minor as it reduces fringing at scales that are  larger than those used in our fittings.

\end{document}